\documentclass[12pt]{article}

\usepackage{amssymb,amsmath,amsfonts,amssymb}
\usepackage{graphics,graphicx,color,hhline}
\usepackage{bbm}
\usepackage{stmaryrd}
 \usepackage{mathtools}
\usepackage{authblk}
\usepackage{euscript}

\def\@abssec#1{\vspace{.05in}\footnotesize \parindent .2in 
{\bf #1. }\ignorespaces} 

\graphicspath{{/EPSF/}{Figures/}}   

\def \Rm {\mathbb R}

\newcommand{\eps}{\varepsilon}
\newcommand{\E}{\mathbb E}

\newcommand{\be}{\begin{equation}}
\newcommand{\ee}{\end{equation}}
\newcommand{\bea}{\begin{eqnarray}}
\newcommand{\eea}{\end{eqnarray}}
\newcommand{\bee}{\begin{eqnarray*}}
\newcommand{\eee}{\end{eqnarray*}}

\newcommand{\bu}{\mathbf u} \newcommand{\bv}{\mathbf v}
 
\newcommand{\br}{\mathbf r}

\newcommand{\bx}{\mathbf x} \newcommand{\by}{\mathbf y}
\newcommand{\bz}{\mathbf z}

\def\fref#1{{\rm (\ref{#1})}}

\newcommand{\calC}{\mathcal C}
\newcommand{\calH}{\mathcal H}

\newcommand{\calR}{\mathcal R}

\newcommand{\calK}{\mathcal K}
\newcommand{\calG}{\mathcal G}

\newcommand{\calT}{\mathcal T}
\newcommand{\bzero}{\mathbf 0}

\newcommand{\cout}[1]{}

\title{Speckle imaging with blind source separation and total variation deconvolution}
 
 \author[1,2]{Randy Bartels\footnote{rbartels@morgridge.org}}
 \author[3]{Olivier Pinaud\footnote{olivier.pinaud@.colostate.edu} }
 \author[4]{Maxine Varughese\footnote{Maxine.Xiu@colostate.edu}}
 \affil[1]{Morgridge Institute for Research, Madison, WI 53715 USA}
 \affil[2]{Biomedical Engineering Department, University of Wisconsin--Madison, Madison, WI 53715 USA}
 \affil[3]{Department of Mathematics, Colorado State University\\ Fort Collins CO, 80523}
\affil[4]{Department of Electrical and Computer Engineering, Colorado State University\\ Fort Collins CO, 80523}
\begin{document}
\maketitle
\begin{abstract}
  This work is concerned with optical imaging in strongly diffusive environments. We consider a typical setting in optical coherence tomography where a sample is probed by a collection of wavefields produced by a laser and propagating through a microscope. We operate in a scenario where the illuminations are in a speckle regime, namely fully randomized. This occurs when the light propagates deep in highly heterogeneous media. State-of-the-art coherent techniques are based on the ballistic part of the wavefield, that is the fraction of the wave that propagates freely and decays exponentially fast. In a speckle regime, the ballistic field is negligible compared to the scattered field, which precludes the use of coherent methods and different approaches are needed. We propose a strategy based on blind source separation and total variation deconvolution to obtain images with diffraction-limited resolution. The source separation allows us to isolate the fields diffused by the different scatterers to be imaged, while the deconvolution exploits the speckle memory effect to estimate the distance between these scatterers. Our method is validated with numerical simulations and is shown to be effective not only for imaging discrete scatterers, but also continuous objects. 
   \end{abstract}

   \section{Introduction}
   We address in this work the optical imaging of objects in strongly diffusive environments. As light propagates in a heterogeneous medium, a fraction of the light behaves as if the medium were homogeneous, while the remaining part is scattered by the inhomogeneities. The first contribution is referred to as the \textit{ballistic part} of the wave, and is exponentially damped as the light travels deeper into the medium. \cite{dunsby2003techniques, rotter2017light} Many state-of-the-art imaging techniques, such as confocal microscopy, \cite{sheppard1978depth, indebetouw1990distortion, anderson1994microscope, sun1994broad} nonlinear microscopy, \cite{beaurepaire2001ultra, helmchen2005deep, masihzadeh2009enhanced, field2016superresolved, fantuzzi2023wide, xu2024multiphoton} optical coherence tomography (OCT), \cite{huang1991optical, fercher1996optical, bouma2022optical} SMART-OCT \cite{smartOCT} and closed-loop accumulation of single scattering (CLASS) \cite{CLASS, yoon2020deep}, utilize the ballistic part only. The use of ballistic light with high numerical aperture (NA) objective lenses ensures that the images obtained from these techniques reach an imaging resolution that is close to the diffraction limit. In situations where the objects to be imaged are buried deep into the medium, however, the ballistic field is extremely small compared to the scattered field and it becomes a challenge to extract the ballistic wave from the total signal. This is achieved in SMART-OCT and CLASS by exploiting coherence effects that are present in the ballistic wave and not in the scattered wave.

   In biological tissues, ballistic methods perform well up to depths ranging from about tens of microns up to a few millimeters, depending on the imaging modality. \cite{badon2017multiple, ntziachristos2010going} Beyond such depths, the ballistic light is unexploitable and a different approach is needed. Such imaging strategies must rely on the scattered field since it overwhelmingly dominates the measured ballistic signal. The crucial difficulty is then to leverage the information carried by the scattered part, \cite{arridge1999optical, o2012diffuse, torabzadeh2017compressed, torabzadeh2019hyperspectral, carminati2021principles} while this is fairly direct with the ballistic part once it is extracted. For example, wave-imaging techniques like ultrasound and seismic imaging \cite{aubry2009random, blondel2018matrix}  rely on methods that isolate the single-scattered field from the multiply scattered field to obtain high resolution optical image information in the first-Born approximation. 
   It is also unclear what imaging resolution can be achieved in such a situation. While deeper imaging is possible with diffuse optical techniques, \cite{lyons2019computational, applegate2020recent} imaging resolution is limited to a transverse spatial resolution on the order of the depth of imaging within the tissue, \cite{moon1993resolution} and consequentially other models for image estimation are required to recover diffraction-limited imaging resolution performance. 

   We propose in this work a strategy allowing us to image objects with diffraction-limited resolution in a configuration where the scattered field is in a \textit{speckle regime}, namely the field can be modeled by a fully randomized wave. There is no ballistic part in such a regime. Our method is based on the framework of the \textit{reflection matrix}, see e.g. \cite{GiganTR}, and our imaging configuration follows the standard experimental setting of OCT: the  sample to be imaged is placed at one conjugate plane of a microscope illuminated by a laser, and a camera placed at the other conjugate plane collects the reflected intensity. Interferometry techniques then allow one to recover the whole field and not just its intensity. \cite{masihzadeh2010label, winters2012measurement, smith2013submillisecond, young2015pragmatic,hu2020harmonic, farah2024synthetic} In addition, a time gating procedure makes it possible to select a particular depth in the sample, with an axial resolution inversely proportional to the bandwidth of the laser (the axial resolution is typically on the order of a few micrometers, with a laser central wavelength between 500 nanometers and 1 micrometer). \cite{dubois2004ultrahigh, murray2023aberration} The measured field is therefore a combination of scattered fields emanating from the same slab at a given depth, ignoring most of the signal scattered in slabs with different travel times.

   In this strategy, the sample is probed by a series of random illuminations, and the associated measured fields are stored in a matrix $\calR$ (after flattening of the 2D fields into a vector), where a column of $\calR$ corresponds to a measured field for one illumination. Our method is inspired by that of \cite{GiganNMF} for fluorescence microscopy, and is conceptually quite similar to the decomposition of the time reversal operator (DORT) method \cite{DORT1,DORT2}, even though the latter operates in a completely different regime where propagation is mostly ballistic. There are two steps. First, the fields emanating from different scatterers are separated, which gives us access to the Green's functions associated with different point scatterers. This separation is the most difficult part. In the DORT method, a standard singular value decomposition (SVD) is enough, provided the scatterers are sufficiently far apart and have sufficiently different intensities. The scatterers separation assumption ensures the associated Green's functions (actually simple plane waves in DORT) are nearly orthogonal, but also limits the applicability of the method. To achieve higher resolution and expand the class of objects that can be imaged, we need a better separation method. The problem falls into the category of \textit{blind source separation (BSS)} problems and is solved using \textit{independent component analysis (ICA)} techniques. Not only do ICA methods improve on the separation provided by the SVD, but they also separate fields when the SVD completely fails to, which occurs when e.g. the scatterers are too close to each other. In particular, there is a resonance effect, which we refer to as the \textit{speckle bond resonance} in reference to the chemical bond resonance in the physics literature, \cite{teller1937crossing, von1990generalization} that prevents the SVD separation of two scatterers with comparable intensities even though the associated Green's functions are nearly orthogonal. The ICA method we are using is immune to this resonance effect.

   The second step exploits these individual Green's functions to locate the scatterers. In DORT, this scatterer localization is immediately done with an inverse Fourier transform applied to the conjugate of the SVD singular vectors, since the Green's functions need to be phase conjugated to back-propagate to the scatterer source. In the case of the imaging configuration presented here, the Green's functions lie in a plane wave basis and the image point is obtained directly through inverse Fourier transform. This is not possible in our speckle regime since the Green's functions are fully random. As in \cite{GiganNMF}, we then exploit the so-called \textit{memory effect}, namely the shift invariance of the speckle: when two scatterers are sufficiently close, the corresponding measured fields are shifted versions of each other, and estimating the shift gives the distance between the scatterers. The shift estimation is done with a standard total variation deconvolution: the total variation regularization ensures that the convolution kernel (or its derivative depending on the chosen functional) is sparse, and the optimization problem then essentially boils down to locating the support of the kernel or of its derivative.

   The resolution obtained in our method is dictated by the correlation length of the unknown speckle that illuminates the sample. This length depends on the underlying heterogeneous medium of propagation and on the experimental apparatus. In the setting considered in this work, the correlation length is equal to the usual diffraction-limited length $\lambda/2 \, \mathrm{NA}$, for $\lambda$ the central wavelength of the laser and $\mathrm{NA}$ the NA of the microscope objective. We are then able to image objects with diffraction-limited resolution in a speckle regime. The method is robust in that it does not depend on any particular model for wave propagation or on a special type of illuminating pattern set, as long as the incoming field decorrelates at a given length and the backscattered field enjoys the memory effect.



   Our current method is limited by the fact that common BSS methods require the sources to be separated to not be Gaussian random variables. Since the fields scattered by the objects are in a speckle regime and are well modeled by Gaussian random fields, the use of off-the-shelf ICA algorithms such as FastICA or RobustICA is precluded in a typical OCT-like imaging procedure where signals are generated through linear optical scattering. This constraint leads us to consider the setting of \textit{second harmonic generation (SHG) microscopy}, a powerful approach in its own right \cite{aghigh2023second} that for us here presents the advantage of solving the Gaussianity issue, since the fields involved in the separation are squares of the speckle fields and are therefore not Gaussian. This strategy can be readily extended to other coherent nonlinear scattering mechanics. \cite{squier1998third, masihzadeh2009enhanced, hoover2012eliminating, masihzadeh2015third, heuke2019coherent, heuke2020spatial, smith2024low} We are currently investigating a method for separating Gaussian fields that will be applied in a linear scattering context.

   In what follows, we present our measurement setting and model as well as imaging strategy in Section \ref{sec:models}; analyze the SVD of the reflection matrix and introduce the blind source separation problem in Section \ref{sec:SVD}; devote Section \ref{sec:TV} to the total variation deconvolution, and Section \ref{sec:simu} to our model for wave propagation and numerical results. Finally, in an Appendix we compute analytically the memory length and the correlation length of the speckle. 
   
  \paragraph{Acknowledgment.} This work was supported by NSF grant DMS-2404785 and grants 2023-336437 and 2024-337798 from the Chan Zuckerberg Initiative DAF, an advised fund of Silicon Valley Community Foundation.

    \section{Measurements model and imaging strategy} \label{sec:models}

    \subsection{Measurement setting}

\begin{figure}[h!]
\begin{center} 
  \includegraphics[scale=1.2]{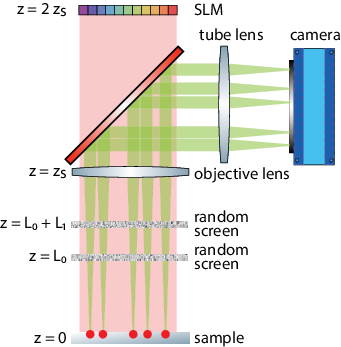}
  \end{center}
  \caption{Measurement setting}
  \label{fig1}
\end{figure}

   Our measurement setting is that of a standard widefield OCT experiment, see e.g. \cite{AubryDO} for a complete description, or \cite{MurraySHG, farah2024synthetic} for a setting specialized to SHG or THG (third harmonic generation), respectively. We do not need here all the experimental details and for our purpose the setting sketched in Figure \ref{fig1} is sufficient: the laser axis is along the $z$ direction, the microscope lens is at height $z=z_s$ from some reference, and there is a spatial light modulator (SLM) located at height $2 z_s$. The latter allows one to impose an arbitrary spatially varying phase to the wavefield at $z=2 z_s$. We set $z_s$ to be the focal distance of the microscope objective; see \cite{mertz2019introduction} for a precise definition (within the Fresnel approximation, the 2D field at $z=0$ is then a scaled 2D Fourier transform of the 2D field in the plane at $z=2z_s$). The object to be imaged lies in the plane at $z=0$, being illuminated by the field that has passed through the two random phase plates and has been randomized to produce a speckle field that obeys circular Gaussian statistics. This speckle field drives scattering in the object, producing a backscattered light collected by the microscope. Linear scattering would produce an object field that replicates the Gaussian statistics of the illumination field, for which our method would fail to produce an image. We consider the SHG scattered field which is radiated at twice the incident fundamental frequency so that the image field is isolated in the backward direction with a dichroic beamsplitter that reflects the second harmonic field and transmits the fundamental field. The backpropagating field again passes through the random phase screens, scrambling the phase of the image field. The reflected field is brought to an image on a camera where measurements are made. In the measurements, a coherent reference beam is introduced to interfere with the image field to perform a holographic measurement as illustrated in Figure \ref{fig1}. The SHG field is estimated from the measured interferogram intensity recorded on the camera. \cite{masihzadeh2010label, smith2013submillisecond} Mathematically, this is expressed by assuming that the camera is located at a plane of height $z=2z_s$. With such a setting, we will probe a sample around the height $z=0$.

   The central quantity is the reflection matrix $\calR\equiv (\calR_{ij})_{i=1,\cdots,N_p,\; j=1,\cdots,N_r}$, where $i$ indexes the pixels on the camera where the SHG field is measured, and where $j$ is the label of one illumination. We assume the field is measured at $N_p$ pixels and that there are $N_r$ illuminations. The construction of the reflection matrix is illustrated in Figure \ref{fig1}. For each illumination, a speckled illumination field is generated that is statistically independent of all of the other illumination fields thanks to application of a random phase on the SLM. Each field estimated from the SHG hologram is flattened from a 2D $\sqrt{N_p} \times \sqrt{N_p}$ matrix to a column vector of length $N_p$. The reflection matrix is assembled from the columns for each of the $N_r$ illumination fields.
   
       \subsection{Model for the reflection matrix}
The structure of $\calR$ is as follows. We start by expliciting its harmonic components at angular frequency $\omega$. The fundamental field is denoted by $U_\omega^{(1)}$ and the SHG field by $U_\omega^{(2)}$. They satisfy the coupled Helmholtz equations below
      \bee
   \Delta U_\omega^{(1)}+k^2(\br,\omega)U_\omega^{(1)}&=&S(\br,\omega), \qquad \br\in \Rm^3\\
   \Delta U_\omega^{(2)}+k^2(\br,\omega)U_\omega^{(2)}&=&-V(\br,\omega) U_\omega^{(1)}(\br) \star_\omega U_\omega^{(1)}(\br), \qquad \br\in \Rm^3
   \eee
   where $\star_\omega$ denotes convolution w.r.t. the variable $\omega$, $S(\br, \omega)$ is a source generating the laser beam, and where
   $$
   V(\br,\omega)=\frac{\omega^2}{2 c_0^2} \chi^{(2)}(\br), \qquad k^2(\br,\omega)=\frac{\omega^2}{c_0^2}n^2(\br,\omega).
   $$
   Above, $n(\br,\omega)$ is the refractive index, $c_0$ the background velocity (supposed to be constant) and $\chi^{(2)}(\br)$ is the second order nonlinear susceptibility (that is assumed to be scalar) of the SHG active scatterers. The fundamental field is centered about the frequency $\omega_0$, with the spectrum extended over a bandwidth $\Delta \omega_1$, and similarly the second harmonic field is centered at twice the fundamental at $2 \omega_0$ with a bandwidth of $\Delta \omega_2$. Note that we are operating under the undepleted pump approximation where the fundamental is not modified by the SHG field, and that the Helmholtz equations are equipped with Sommerfeld radiation conditions. 
   
   We rewrite the fundamental field $U_\omega^{(1)}$ as   $U_\omega^{(1)}(\br)= e^{i \omega (2 z_s-z)/ c_0}\textsf{U}_{\omega}^{(1)}(\br) \, A(\omega)$ where $\br=(\bx,z)$ for $\bx \in \Rm^2$, and $A(\omega)$ is the spectral field of the fundamental pulse, which is centered at $\omega_0$ with bandwidth $\Delta \omega_1$. Assuming propagation is forward-directed and that the laser beam has a narrow bandwidth (in practice $\Delta \omega_1 / \omega_0$ is about a few percents), the envelope $\textsf{U}_{\omega}^{(1)}$ is slowly varying with respect to $\omega$ and we have $\textsf{U}_{\omega}^{(1)} \simeq \textsf{U}_{\omega_0}^{(1)}$. The autocorrelation in the source term for the SHG wave equation is then simplified to
   $$ U_\omega^{(1)}(\br) \star_\omega U_\omega^{(1)}(\br) \simeq e^{i \omega (2z_s-z)/ c_0} (\textsf{U}_{\omega_0}^{(1)}(\br))^2 \, \mathcal{A}(\omega),$$ where $\mathcal{A}(\omega)$ is the autocorrelation of the fundamental pulse spectrum $A(\omega)$, which is hence centered to $2 \omega_0$. 
   
We denote by $\calG_{\omega}$ the Green's function satisfying
   \bee
      \Delta \calG_{\omega}(\br;\br')+k^2(\br,\omega)\calG_{\omega}(\br;\br')&=&-\delta(\br-\br'), \qquad \br,\br'\in \Rm^3
      \eee
      with Sommerfeld radiation conditions. This allows us to express $U^{(2)}_\omega$ as
   $$
    U^{(2)}_\omega(\br')=\int_{\Rm^3} \calG_{\omega}(\br';\br) V(\br,\omega)  e^{i \omega (2z_s-z)/ c_0}  (\textsf{U}_{\omega_0}^{(1)}(\br))^2 \, \mathcal{A}(\omega) d\br.
    $$
We will write further $\calG_{\omega}(\bu,z';\br)$ and $\calG_{\omega}(\bu,z';\bx,z)$ for $\bu,\bx \in \Rm^2$ when the $z$ component of $\br',\br$ needs to be explicitly indicated.  We denote by $j=1,\cdots,N_r$ the label of the illumination and introduce the following notations:
     \begin{align*}
    &  \calH^{j}_\omega(\br)= e^{i \omega (2z_s-z)/ c_0} (\textsf{U}_{\omega_0,j}^{(1)}(\br))^2 \, \mathcal{A}(\omega) \\
      &U^{(2)}_{\omega,j}(\bu) =\int_{\Rm^3} \calG_{\omega}(\bu,2 z_s;\br) V(\br,\omega)  \calH^{j}_\omega(\br) d\br.
   \end{align*}
   The variable $\bu$ is a point on the camera where measurements are made. 
Standard optical holography techniques then give measurements of the form 
\bee
   \calR(\bu, j) =  \int_{\Rm} U^{*}_{\textrm{ref},\omega}(\bu) U^{(2)}_{\omega,j}(\bu)  d \omega
   \eee
   where $U_{\textrm{ref},\omega}$ is a reference field. The latter is spatially uniform in a typical holography configuration so that we may write $U_{\textrm{ref},\omega}(\bu) = \exp(i \omega T) \, \mathcal{A}_{\textrm{ref}}(\omega)$, where $T$ is an experimental parameter and $\mathcal{A}_{\textrm{ref}}(\omega)$ is the reference field spectrum centered at $2\omega_0$. The time $T$ is set to be the travel time back and forth between the conjugate planes of the microscope objective, namely the planes at $z=0$ and at $z=2z_s$. We then have $T=T_s=4z_s/c_0$ and the reflection matrix $\calR(\bu,j)$ takes the form
\bee
   \calR(\bu,j)= \int_{\Rm} \int_{\Rm^3} e^{-i \omega T_s}  \mathcal{A}_{\textrm{ref}}^*(\omega) \calG_{\omega}(\bu,2 z_s;\br) V(\br,\omega)  \calH^{j}_\omega(\br) d\br d \omega.
\eee
      
We will need the following notations:
\begin{align*}
  &\calG_{\omega}(\bu,z;\bx',z')=e^{i \omega |z-z'|/c_0 }G_{\omega}(\bu,z;\bx',z')\\
  &H^j_{\omega_0}(\bx,z)= (\textsf{U}_{\omega_0}^{(1)}(\br))^2.
\end{align*}
The function $\hat{s}(\omega) = \mathcal{A}_{\textrm{ref}}^*(\omega)  \mathcal{A}(\omega) $ below denotes the signal-reference cross spectral density centered at $2 \omega_0$ with bandwidth $\Delta \omega$. Under forward-directed propagation and narrow bandwidth assumptions, the field $G_{\omega}$ varies slowly w.r.t. $\omega$, so that
$\hat{s}(\omega) G_{\omega}  \simeq \hat{s}(\omega) G_{2\omega_0}$ and $\hat{s}(\omega) V(\bx,z',\omega)  \simeq \hat{s}(\omega) V(\bx,z',2\omega_0)$. Let $l_{\rm{coh}}=c_0/\Delta \omega$ be the cross-coherence length of the SHG signal and reference fields and write $s(t)= s_0(-c_0 t/4l_{\rm{coh}})/2 \pi$ for the inverse Fourier transform of $\hat s$. Then,
    \bee
    \calR(\bu,j) &=& \int_{\Rm} \int_{\Rm^3}  e^{i \omega ((4 z_s-2z')/c_0-T_s)} G_{2\omega}(\bu,2 z_s;\bx,z') V(\bx,z',\omega)  H^j_{\omega_0}(\bx,z') \hat s(\omega) d\bx dz' d\omega \\
    &\simeq &\int_{\Rm^3}  G_{2\omega_0}(\bu,2 z_s;\bx,z') V(\bx,z',2\omega_0)  H^j_{\omega_0}(\bx,z')s_0(z'/l_{\rm{coh}}) d\bx dz'. 
    \eee
    
Assuming that both $G_{2\omega_0}$ and $H^j_{2\omega_0}$ vary slowly w.r.t. $z$ across a coherence length $l_{\rm{coh}}$, we find, if e.g. $s_0(x) = 1$ for $|x| \leq 1/2$ and $s_0(x)=0$ for $|x|>1/2$, 
    \bee
    \calR(\bu,j) 
    &\simeq &  \int_{\Rm^2} G_{2\omega_0}(\bu,2 z_s;\bx,0)    V_s(\bx)  H^j_{\omega_0}(\bx,0) d\bx 
    \eee
    where
    $$
    V_s(\bx)=\int_{-l_{\rm{coh}}/2}^{l_{\rm{coh}}/2} V(\bx,z,2\omega_0) dz.
    $$
The quantity $\calR(\bu,j)$ is in practice an excellent approximation of the field that is measured and it incorporates the time gating mechanism of OCT---that is, the measured signal consists mostly of the field backscattered by a slab of width $l_{\rm{coh}}$ around $z=0$. The length $l_{\rm{coh}}$ gives the axial resolution of the method and it is clear from the definitions of $V_s$ and $\calR(\bu,j) $ that it is not possible to extract the variations of $\chi^{(2)}$ within one coherence length.

Simplifying notation so that $G_{2\omega_0}(\bu;\bx)=G_{2\omega_0}(\bu,2 z_s;\bx,0) $  and $H^j_{\omega_0}(\bx)=H^j_{\omega_0}(\bx,0)$, the reflection matrix has the final form, for $i=1,\cdots,N_p$, $j=1,\cdots,N_r$,
    $$
     \calR_{ij}=\calR(\bu_i,j)=\int_{\Rm^2} G_{2\omega_0}(\bu_i;\bx)    V_s(\bx)  H^j_{\omega_0}(\bx) d\bx.
     $$
     The $N_r$ illuminations are obtained by randomly choosing $N_r$ random phases on the SLM. See Section \ref{sec:simu} for more details.


            \subsection{Imaging strategy}

     Knowing $\calR$, the goal is to reconstruct $V_s$, or at least its support, which is the object that we seek to image. In the DORT method, this is accomplished under the key assumption that $G_{2\omega_0}$ and $H^j_{\omega_0}$, or at least their physical model, are known. This occurs for instance when the light propagates in free space, and depending on the experimental setting, $G_{2\omega_0}$ and $H^j_{\omega_0}$ can either be plane waves, approximations of Dirac deltas, or the Green's function for the case of point scatterers. When the scatterers are sufficiently well separated and have sufficiently different intensities, orthogonality in $G_{2\omega_0}$ and $H^j_{\omega_0}$ can be exploited and an SVD performs well in separating the fields $G_{2\omega_0}(\bu;\bx_j)$ at the different locations $\bx_j$ of the scatterers. It suffices then to ``invert'' $G_{2\omega_0}$ to find the scatterers.

     We assume here that the light propagates in a highly diffusive environment, and as a consequence neither $G_{2\omega_0}$ nor $H^j_{\omega_0}$, nor their forms, are known, as we do not have knowledge of the scattering potential. This makes the imaging problem considerably more difficult. First, the SVD does not separate scatterers as well or at all the fields $G_{2\omega_0}(\bu;\bx_j)$, and further, the action of $G_{2\omega_0}$ cannot be removed since it is not known. We will use the following two properties of the fields $G_{2\omega_0}$ and $H^j_{\omega_0}$ to solve the problem. As detailed in Section \ref{sec:SVD}, after propagation in a strongly inhomogeneous medium, the fields tend to be in a speckle regime, such that at each spatial point, the field is a complex circular Gaussian random variable (i.e. of the form $X+iY$, with $\E\{X\}=\E\{Y\}=\E\{X Y\}=0$, $\E\{X^2\}=\E\{Y^2\}$). The exact form of the spatial correlation function is not important in our method; the only relevant parameter is the spatial correlation length of the Gaussian field. We distinguish the correlation length $\ell_{s,\rm{in}}$ of the incoming field from that of the outgoing SHG field, denoted $\ell_{s,\rm{out}}$. Both $\ell_{s,\rm{in}}$ and  $\ell_{s,\rm{out}}$ are relevant in the SVD separation, while $\ell_{s,\rm{in}}$ is more key to the ICA. The latter also eventually gives the (transverse) resolution. We evaluate analytically $\ell_{s,\rm{in}}$ and  $\ell_{s,\rm{out}}$ in Appendix \ref{app:corr} for the propagation model we consider in the simulations, and find that both are of order of the central wavelength in typical experimental settings. Moreover, since the field $H^j_{\omega_0}$ is the square of a Gaussian field in a SHG context, it is not itself Gaussian. The SVD separation can then be improved using ICA algorithms that are applicable to non-Gaussian fields. 

     The second property we use, detailed in Section \ref{sec:TV}, is that if two speckle fields originate from two scatterers that are separated by a distance less than the memory length $\ell_{\rm{me}}$, they are approximately shifted versions of each other.  To estimate the distance between the scatterers, it suffices then to estimate the shift, which is done with a total variation deconvolution algorithm.

     \section{SVD and blind source separation} \label{sec:SVD}

     We start by analyzing the SVD and then turn to the ICA.

    We first equivalently discretize the integral in $\calR$ using a grid $(\bx_j)_{j=1,\cdots,N}$ with $N$ points or assume $\chi^{(2)}$ consists of $N$ point scatterers located at $(\bx_j)_{j=1,\cdots,N}$. The reflection matrix then reads ($^T$ denotes transposition)
   
$$
     \calR_{ij}=\sum_{\ell=1}^{N}G_{2\omega_0}(\bu_i;\bx_\ell)    V_s(\bx_\ell)  H^j_{\omega_0}(\bx_\ell) = G \rho H^T
     $$
     where $G$ is a $N_p \times N$ matrix with components $G_{ij}=G_{2\omega_0}(\bu_i;\bx_j) $, $\rho$ is a diagonal $N \times N$ matrix with diagonal entries $\rho_{jj}= V_s(\bx_j) $, and $H$ is a $N_r \times N$ matrix with components $H_{j \ell}=H^j_{\omega_0}(\bx_\ell)$.

      Given $\calR$, the main objective of the section is to individually extract the columns of $G$. To do so, we first compute the SVD of $\calR$ that results in $\calR=U \Sigma V^*$, where $^*$ denotes conjugate transposition. We then need to relate $U,V$ with $G$ and $H$ as below.
     
     \subsection{The singular vectors}
     \paragraph{Relation between $U$ and $G,H$.} We recall that the matrix $U$ is unitary and stores the eigenvectors of the matrix $\calR \calR^*$, which reads 
     $$
     \calR \calR^*= G \rho H^T \bar{H} \rho^* G^*,
     $$
     where $\bar{H}$ is the complex conjugate of $H$. Gram-Schmidt orthogonalization allows us to write
$$
G=Q_G R_G, \qquad \bar H=Q_H R_H,
$$
where $Q_G$ and $Q_H$ have $N$ orthonormal columns and $R_G$, $R_H$ are $N \times N$ upper triangular matrices. We next diagonalize the Hermitian matrix $R_G \rho H^T \bar{H} \rho^* R_G^*$ as $R_G \rho H^T \bar{H} \rho^* R_G^* =S_U D_U S_U^*$, where $S_U$ is unitary and $D_U$ diagonal. Then $\calR \calR^*= Q_G S_U D_U S_U^*  Q_G^*$. This shows that the unitary matrix $Q_G S_U$ contains the eigenvectors of $\calR \calR^*$. Since those are known up to permutations and multiplications by a complex factor with modulus one, we can write
     \be \label{defU}
     U= Q_G S_U U_U P_U=G R_G^{-1} S_U U_U P_U
     \ee
     where $U_U$ is a diagonal unitary matrix and $P_U$ a permutation matrix. Hence, the columns of $U$ are linear combinations of the columns of $G$, namely of the vectors $(G_{2\omega_0}(\bu_i;\bx_j))_{i=1,\cdots,N_p}$.

     Assume for the moment that the columns of both $G$ and $H^T$ are orthogonal. Then $H^T \bar{H}$ is diagonal and $\calR \calR^*= G D G^*$ for a diagonal matrix $D$. Hence, the columns of $U$ are proportional to those of $G$ (up to permutations) and the SVD perfectly separates the fields emitted by the scatterers. In practice, orthogonality is only approximate, and this gives rise to an interesting resonance phenomenon that prevents separation with the SVD. This can be explained with a simple system with two scatterers as follows.

     Set $N=2$, and write
     $$
     H^T \bar{H}=
     \begin{pmatrix}
       1 & \eta\\
       \eta & 1
     \end{pmatrix}, \qquad \rho=
     \begin{pmatrix}
      \rho_1 & 0\\
       0 & \rho_2
     \end{pmatrix}, \qquad R_G=
     \begin{pmatrix}
      1 & \eps \\
       0 & 1
     \end{pmatrix}.
     $$
     Above, we assumed without lack of generality that the columns of $H^T$ and $G$ are normalized to one. The parameters $\eta$ and $\eps$ measure how orthogonal the columns of $G$ and $H^T$ are, and are set to be real-valued for simplicity, as well as $\rho_1$ and $\rho_2$. Then,
     $$
    M=R_G \rho H^T \bar{H} \rho^* R_G^*   =S_UD_US_U^*=\begin{pmatrix}
       a_1 & b\\
      b &a_2
     \end{pmatrix}
     $$
     where
     \begin{align*}
       &a_1=\rho_1^2+2 \eps \eta \rho_1\rho_2 + \eps^2 \rho_2^2, \qquad a_2=\rho_2^2\\
 &      b=\eta \rho_1 \rho_2+\eps\rho_2^2.
     \end{align*}
The eigenvalues of $M$ are $\frac{1}{2}(a_1+a_2\pm \sqrt{(a_1-a_2)^2+4 b^2})$ and the eigenvectors (not normalized) are
     $$
     v_\pm=\begin{pmatrix} 1 \\
       (\delta a\pm \sqrt{(\delta a)^2+4 b^2})/2b
       \end{pmatrix}, \qquad \delta a=a_2-a_1.
       $$
       The eigenvectors are stored in $S_U$. Owing to expression \fref{defU}, the columns of $U$ are approximately proportional to those of $G$ (up to permutations) provided that: first, $R_G^{-1}$ is nearly diagonal, that is when $\eps \ll 1$, namely the columns of $G$ are nearly orthogonal, and second, if $S_U$ is nearly diagonal, up to permutations. The latter holds when the second component of $v_\pm$ is either much larger or much smaller than the first one, i.e. 1. This occurs when $2|b| \ll |\delta a|$, in which case, say when $\delta a<0 $, we have $v_+\simeq (1,-b/\delta a)$ and $v_- \simeq (1,\delta a /b)$. To leading order, we have $\delta a\simeq \rho_2^2-\rho_1^2$ and this successful separation scenario corresponds to the case where the columns of $G$ and $H^T$ are nearly orthogonal and the scatterers have sufficiently different intensities.

       On the contrary, when $2 |b|$ and $|\delta a|$ are comparable, independently of how small the coupling term $b$ is, the components of $v_\pm$ have similar size: set for instance $\delta a =-2 b $, then $v_\pm=(1, -1\pm \sqrt{2})$. In such a situation, the SVD does not separate the columns of $G$ and $H^T$, even though the fields are nearly orthogonal. This phenomenon, that we call the \textit{speckle bond resonance}, is quite general: the normalized eigenvectors of $M$ obtained when $b=0$, i.e. $(1,0)$ and $(0,1)$, become significantly coupled for $b \neq 0$ whenever the condition $2|b| \ll |\delta a|$ is not verified. In particular, when the scatterers have similar intensities, we have $\delta a \simeq 0$ and separation will be poor. 
       
     
     \paragraph{Relation between $V$ and $G,H$.} The analysis is similar to the previous one and we find
     \be \label{defVA}
     V^*=A H^T, \qquad A=(R_H^{-1} S_H U_H P_H)^*
     \ee
where $U_H$ is a diagonal unitary matrix and $P_H$ a permutation matrix, and $
     R_H \rho^* G^* G \rho R_H^*=S_H D_H S_H^*
     $ with $S_H$ unitary and $D_H$ diagonal. The columns of $V^*$ are therefore linear combinations of those of $H^T$, and are proportional to those of $H^T$ up to permutations when both the columns of $G$ and $H^T$ are orthogonal. 

     As before, the speckle bond resonance holds and the SVD alone cannot separate the fields well enough for imaging. We then improve on the separation by using ICA, which provides us with an estimate of $A^{-1}$ (up to permutations and rescaling of the columns).

          \subsection{Blind source separation}

          We first describe the general setting of BSS and then particularize to our imaging problem. Consider the linear system $x=As$, where $A$ is a (deterministic) complex-valued $N\times N$ invertible matrix, $s \equiv (s_i(\omega))_{i=1,\cdots,N}$ is a complex-valued random (column) vector of length $N$ ($\omega$ denotes in this section an element of a probability space that does not need to be specified). The matrix $A$ and the source $s$ are unknown, and the goal is to reconstruct $s$ from knowledge of $x(\omega)$ only. This is done using ICA.

          The main idea behind standard ICA is to maximize non-Gaussianity: suppose for the moment that the components $s_i$ are independent with same distributions; according to the central limit theorem, a linear combination of the $s_i$ will be closer to a Gaussian distribution than the $s_i$ individually. To separate the $s_i$, one then optimizes some criterion that characterizes Gaussianity. One such criterion used in the FastICA \cite{fastICA} and RobustICA \cite{zarzoso2009robust} algorithms is based on the kurtosis, and we use here the definition given in RobustICA: let $w$ be a column vector with norm 1 and denote by $w^*$ its conjugate transpose. Given $x=As$ and writing $y=w^* x$, the kurtosis considered in RobustICA has the form
$$
K(w)=\frac{\E \{|y|^4\}- 2 (\E\{|y|^2\})^2-|\E\{y^2\}|^2}{(\E\{|y|^2\})^2}.
$$
This expression is more general than the one in FastICA which only applies when the $s_i$ are independent, resulting in $\E\{y^2\}=0$. This is not verified in our imaging setting since the $s_i$ are not independent. 

When $x$ is pre-whitened (which is the case for us since $x$ originates from an SVD as we have $x\equiv V^*$), we have $\E \{xx^*\}=I$, for $I$ the identity matrix. With the normalization condition $w^* w=1$, we find
$$\E\{ yy^*\}=\E\{ w^* x x^* w\}=w^* w=1,$$
and $K$ becomes
$$
K(w)=\E \{|y|^4\}-2-|\E\{y^2\}|^2.
$$
When $y$ is a complex Gaussian random variable, its kurtosis is zero. In practice, $s$ is an $N \times N_r$ matrix, where $N_r$ is the number of realizations of the random variables, and only empirical averages are available. The expectation $\E\{(\cdot)\}$ has to then be replaced by 
$\frac{1}{N_r}\sum_{\ell=1}^{N_r} (\cdot)
$. We will keep the notation $\E$ for simplicity, keeping in mind it is an empirical average in our setting. The ICA aims at computing an estimate of $A^{-1}$ (up to arbitrary permutations and rescaling of the columns). In our imaging setting, we have $x=V^*$, $s=H^T$, and $A$ has the expression given in \fref{defVA}. When the $s_i$ are mutually independent, when at most one of the $s_i$ is Gaussian, and when $\Re\{s_i\}$ and $\Im\{s_i\}$ are independent, it is shown in \cite{fastICA} that the columns of $A^{-1}$ are among the critical points of $K(w)$ under appropriate constraints.

In our case, the $s_i$ are in general not independent since the fields associated with nearby scatterers are correlated. We show in a companion paper \cite{ICA-BP} that the independence assumption can be relaxed to a condition on higher moments of $s$, which is somewhat expected since $K$ only involves fourth order moments. We establish that such a condition is approximately verified by the columns of $H^T$, which in turn implies that the columns of $A^{-1}$ are approximately critical points of $K$. More precisely, we obtain decoupling conditions of the form
\be \label{decoup}
\E\{s_i s_j |s_k|^2\} \simeq 0
\ee
with variants of this relation. The $i$ row of $s$ has entries $H^j_{\omega_0}(\bx_i)=(\textsf{U}^{(1)}_{\omega_0}(\bx_i))^2$, $i=1,\cdots,N$, $j=1,\cdots,N_r$, and since the fundamental field $\textsf{U}_{\omega_0}^{(1)}$ is in a speckle regime, the rows of $s$ are approximately given by realizations of squares of complex circular Gaussian random variables and equation \fref{decoup} involves moments of order 8 of $\textsf{U}_{\omega_0}^{(1)}$. We have then $\E\{s_i\}\simeq 0$ as well as $\E\{s^2_i\}\simeq 0$, and higher order moments can be evaluated using Isserlis theorem to arrive at
\begin{align*}
\E&\{s_i s_j^* |s_k|^2\} \simeq 2 (\E\{\textsf{U}^{(1)}_{\omega_0}(\bx_i)\textsf{U}^{(1)}_{\omega_0}(\bx_j)^*\})^2 (\E\{|\textsf{U}^{(1)}_{\omega_0}(\bx_k)|^2\})^2\\
&+8\E\{\textsf{U}^{(1)}_{\omega_0}(\bx_i)\textsf{U}^{(1)}_{\omega_0}(\bx_k)^*\}\E\{\textsf{U}^{(1)}_{\omega_0}(\bx_j)^* \textsf{U}^{(1)}_{\omega_0}(\bx_k)\}\E\{\textsf{U}^{(1)}_{\omega_0}(\bx_i)\textsf{U}^{(1)}_{\omega_0}(\bx_j)^*\}\E\{|\textsf{U}^{(1)}_{\omega_0}(\bx_k)|^2\}\\
&+8\E\{\textsf{U}^{(1)}_{\omega_0}(\bx_i)\textsf{U}^{(1)}_{\omega_0}(\bx_k)\}\E\{\textsf{U}^{(1)}_{\omega_0}(\bx_j)^*\textsf{U}^{(1)}_{\omega_0}(\bx_k)^*\}\E\{\textsf{U}^{(1)}_{\omega_0}(\bx_i)\textsf{U}^{(1)}_{\omega_0}(\bx_j)^*\}\E\{|\textsf{U}^{(1)}_{\omega_0}(\bx_k)|^2\}.
\end{align*}
The first two terms on the right above are the leading ones, and their behavior follows the correlation $\E\{\textsf{U}^{(1)}_{\omega_0}(\bx)\textsf{U}^{(1)}_{\omega_0}(\by)^*\}$. We obtain in Appendix \ref{app:corr} an expression of the correlation length $\ell_{s,\rm{in}}$ of $\textsf{U}^{(1)}_{\omega_0}$ for our propagation model. Since $\E\{\textsf{U}^{(1)}_{\omega_0}(\bx)\textsf{U}^{(1)}_{\omega_0}(\by)^*\}$ goes to zero as $|\bx-\by| \gg \ell_{s,\rm{in}}$, condition \fref{decoup} is verified when $|\bx_i-\bx_j|\gg \ell_{s,\rm{in}}$ for $i,j=1,\cdots,N$, $i \neq j$, and it is expected that the ICA will separate scatterers at least $\ell_{s,\rm{in}}$ apart.

Running an ICA algorithm with data $V^*$ will then produce an estimate of $A^{-1}$, and consequently of $A$. Once this is done, this allows us to ``remove'' $H^T$ from $\calR= G \rho H^T$ and to estimate $G \rho$ by
$$
G \rho=U \Sigma A \qquad \textrm{up to permutations and rescaling of the columns}.
$$
Since $\rho$ is diagonal, this gives an estimate of the columns of $G$ individually.

From a computational perspective, the separation cannot be perfect but is often sufficiently good for our imaging purposes. Moreover, to the best of our knowledge, there is no theoretical proof of \textit{global} convergence of ICA-type algorithms to the \textit{local} extremizer $A^{-1}$, even when the $s_i$ are independent.  Though, it appears in practice that the RobustICA algorithm converges to a good approximation of $A^{-1}$.

We explain in the next section how to exploit the columns of $G$ to produce an image of the scatterers. For this task, we slightly modify the deconvolution approach of \cite{GiganNMF}.

\section{Total variation deconvolution} \label{sec:TV}
Now that the columns of $G$ have been estimated, we have access to the field emitted by scatterer $j$, that is $G_{2\omega_0}(\bu;\bx_j) $ for $\bu=\bu_i$, $i=1,\cdots,N_p$ and $j=1,\cdots,N$. We then exploit the memory effect to estimate the distance between the scatterers $\bx_i -\bx_j$, $j \neq i$ and obtain an image: suppose $|\bx_i-\bx_j|\leq \ell_{\rm{me}}$, where we recall $\ell_{\rm{me}}$ is the length characterizing the memory effect (an analytical expression of $\ell_{\rm{me}}$ is given in Appendix \ref{app:memory} for our propagation model); then, approximately, $|G_{2\omega_0}(\bu;\bx_j)| \simeq |G_{2\omega_0}(\bu+\alpha_{\rm{me}} (\bx_i-\bx_j);\bx_i)|$ where $\alpha_{\rm{me}}$ is a constant that depends on the propagation model and can be explicitly calculated (note that we take the absolute value of the field to remove phase factors that limit the memory effect). The quantity $|G_{2\omega_0}(\bu+\alpha_{\rm{me}} (\bx_i-\bx_j);\bx_i)|$ can be expressed as $|G_{2\omega_0}(\cdot;\bx_i)| \star \delta_{\alpha_{\rm{me}} (\bx_j-\bx_i)}$ where $\star$ denotes convolution and $\delta_\bx$ the Dirac measure at $\bx$.

Given $G_{2\omega_0}(\bu;\bx_i)$ and $G_{2\omega_0}(\bu;\bx_j)$, and provided $\bx_i$ and $\bx_j$ are sufficiently close, estimating $\bx_i-\bx_j$ then amounts to solving a deconvolution problem to recover the Dirac delta. Given our experimental setting, a natural formulation of the problem is as follows: let $\psi^{(j)}(\bu)=|G_{2\omega_0}(\bu;\bx_j)|$ as well as
$$
\calK(\bu)=\sum_{n=1}^{N_p} a_n \delta(\bu-\bu_n), \qquad 
\qquad J_{ij}(\calK)=\sum_{\ell=1}^{N_p} |\psi^{(j)}(\bu_\ell)-(\calK \star \psi^{(i)})(\bu_\ell)|^2.
$$
The optimization problem is then to find $(a_n,\bu_n)_{n=1,\cdots,N_p }$ that minimizes $J_{ij}(\calK)$. There is a fair amount of ``noise'' in the problem, coming from the facts that the relation $|G_{2\omega_0}(\bu;\bx_j)| \simeq |G_{2\omega_0}(\bu+\alpha_{\rm{me}} (\bx_i-\bx_j);\bx_i)|$ is only an approximation, that the separation is not perfect, that the model for the reflection matrix is also only an approximation, and that there is possibly true experimental noise. A relaxed version of the optimization problem is therefore preferable. Since we are looking for a sparse $\calK$, it is natural to consider an $\ell_1$ regularization of the form
\be \label{BL}
\min_{(a_n,\bu_n)} J_{ij}(\calK)+\tau \sum_{n=1}^{N_p} |a_n|,
\ee
for some regularization parameter $\tau$. This is the classical (discrete) Beurling-Lasso problem, see e.g. \cite{CandesSuper,Duval2013ExactSR} in the context of spikes localization. There is yet another possible formulation: the last term above is the total variation of the measure $\calK$; supposing the measure has a smooth density $f$, a continuous expression of the total variation of $\calK_f$ with density $f$ is $\|\calK_f\|_{TV}=\|\nabla f\|_{L^1}$. The optimization problem based on $J_{ij}(\calK_f)+\tau \|\nabla f\|_{L^1}$ is commonly referred to as a $TV-L^2$ deconvolution or as the Fatemi-Osher-Rudin problem \cite{RudinOsher,RudinOsher2}. The discrete version reads

\be \label{RF}
\min_{f} J_{ij}(\calK_f)+\tau \sum_{n=1}^{N_p} |D f(\bu_n)|,
\ee
where the convolution in $J_{ij}(\calK_f)$ is now a discrete convolution and $D$ denotes the forward finite difference operator discretizing the gradient.

We tested both \fref{BL} and \fref{RF}. Writing the convolution as a matrix-vector product, \fref{BL} can be solved using the FISTA \cite{fista} and GELMA algorithms \cite{Moscoso_2012} that are based on versions of proximal gradient descent. As in \cite{GiganNMF}, we solve \fref{RF} using the method and code of \cite{Chan} based on an augmented Lagrangian and alternating direction methods. It turns out that in practice, \fref{RF} provides much better results in terms of computational cost, signal-to-noise ratio, and resolution. While it is expected that smoother kernels reduce the background noise, a complete understanding of why \fref{RF} performs better than \fref{BL} is still missing. We do not pursue this avenue further here since \fref{RF} provides us with very accurate results that are sufficient for our goals as will be demonstrated in Section \ref{sec:simu}.

In the simulations, we normalize the maximal value of $\psi^{(j)}$ to one (i.e. we replace $\psi^{(j)}$ by $\psi^{(j)}/ \max |\psi^{(j)}|$, where the maximum is taken over the domain where $\psi^{(j)}$ is measured) in order to set one value of $\tau$ that works for all $\psi^{(j)}$ independently of how large their amplitude is.

To construct an image, we slightly modify the procedure and code of \cite{GiganNMF} and add a denoising step: fixing one $i$, $\psi^{(i)}$ is deconvolved with all $\psi^{(j)}$ for $j=1,\cdots,N$, and we denote by $(f_{ij}(\bu_\ell))_{\ell=1,\cdots,N_p}$ the corresponding solutions to \fref{RF}. Let
\be \label{defsig}
\sigma_{ij}=\max_{\bu} |f_{ij}(\bu)|, \qquad \sigma_i=\max_{j=1,\cdots,N,\; j\neq i} \sigma_{ij}.
\ee
Scatterers $j$ that are close to the scatterer with index $i$ have larger $\sigma_{ij}$ and tend to be better estimated than those that are further away and have as a consequence smaller $\sigma_{ij}$. We then set an arbitrary threshold $\sigma$ to remove scatterers that are too far, and a first image $I_i$ is obtained with the formula
\be \label{expI}
I_i=\sum_{j=1}^N |f_{ij}| h_e(\sigma_{ij}- \sigma \sigma_{ii}), \qquad i=1,\cdots,N,
\ee
where $h_e$ is the heaviside function. The procedure is repeated for all $i=1,\cdots,N$, producing a set of images $(I_i)_{i=1,\cdots,N}$. These images are finally merged as in \cite{GiganNMF} by estimating the relative spatial shift between them: let $\bu_{ij}^*$ the point where $\sigma_{ij}$ in \fref{defsig} is achieved, and suppose our first image is $I_1$; the image that will be merged with $I_1$ is the one producing the largest $\sigma_{1j}$ in \fref{defsig}, and let $j_1$ be such that $\sigma_1=\sigma_{1j_1}$.  Image $I_{j_1}$ is then shifted by $\bu^*_{i j_1}$ and added to $I_1$. The procedure is repeated starting from $I_{j_1}$ and so on until all indices are exhausted.

\section{Simulations} \label{sec:simu}

Section \ref{model} is dedicated to the description of our wave propagation model. We study the resolution in Section \ref{reso}, and present imaging results with the SVD separation only in Section \ref{imSVD}. Section \ref{scat} is devoted to the reconstruction of discrete scatterers, and Section \ref{cont} to that of continuous objects. We image scatterers placed beyond the memory window in Section \ref{beyond}, and propose an approach to decrease the number of illuminations in Section \ref{decrease}.

\subsection{Model for wave propagation} \label{model}
We use a standard model in Fourier optics for light propagation that closely matches typical experimental settings and avoids the resolution of a full high frequency 3D wave propagation problem in complex media. The fundamental field $U_{\omega_0}^{(1)}$ is obtained as follows: its value (more precisely its phase) on the SLM plane at $z=2z_{s}$ is controllable and set by the SLM; the speckle regime is obtained by placing two random phase screens at height $L_0$ and $L=L_0+L_1$ ($0<L_0<L<z_s$) between the sample plane and the microscope objective lens (Figure \ref{fig1}). With just one random screen, the memory length $\ell_{\rm{me}}$ is infinite and a second one is added to obtain a finite (controllable) length. These random screens are usually ground glass diffusers acting as rough surfaces.

Propagation between the SLM, through the microscope objective lens, and to the plane at $z=L$ is done using classical Fourier optics; see e.g. \cite{mertz2019introduction}. We write below $U_{\omega_0}^{(1)}(\bx,z)$ for $U_{\omega_0}^{(1)}(\br)$ when $\br=(\bx,z)$. The portion modeled follows the incident wave propagation from the back focal plane (i.e., the pupil plane) of the microscope objective lens situated at $z = 2 \, z_s$ to the front plane of the objective lens, at which our object is located, $z=0$. Denoting the field in the SLM plane at $z=2z_s$ (for consistency with standard Fourier optics notation we write $z_s=f$ for the $f$ microscope focal length) as $E_{\rm{SLM}}$, we have for the field at $z=L$, up to irrelevant multiplicative factors we will systematically ignore in the sequel,
\bee
U_{\omega_0}^{(1)}(\bx,L)&=&  e^{ik (2f -L)} \int_{\Rm^2}e^{i k  L |\by|^2/2f^2} e^{-i k  \bx \cdot \by/f} E_{\rm{SLM}}(\by) P_{\rm{SLM}}(\by) d\by,
\eee
where $k=\omega_0/c_0$ and $P_{\rm{SLM}}$ is the pupil function, here a circular aperture that eventually defines the resolution on the sample plane at $z=0$. Note that we have set $A(\omega)=1$ for the spectral function since it is not relevant here. The above integral is evaluated by a Fast Fourier Transform. The field $U_{\omega_0}^{(1)}(\bx,L)$ then goes through the random phase screen at $z=L$ and propagates to $z=L_0$. We use a simple paraxial (Fresnel) propagator for this, but other paraxial propagators could be equally employed. The field at $z=L_0$ is then
\bee
U_{\omega_0}^{(1)}(\bx,L_0)&=&  e^{ikL_1}\int_{\Rm^2} e^{i k |\bx-\by|^2/2L_1}  e^{i k S_1(\by)} U_{\omega_0}^{(1)}(\by,L) P(\by) d\by,
\eee
where $S_1(\by)=\sigma_1 V_1(\by)$ is the optical path length accumulated by the field propagating through the diffuser, for $V_1$ a mean-zero real-valued random field with unit variance, and where $P$ is the indicator function of the spatial support of the random screen. The above convolution is computed in the Fourier space using the exact expression of the Fourier transform of the kernel $e^{i k |\bx|^2/2L_1}$. The field at $z=0$ is finally obtained by multiplying $U_{\omega}^{(1)}(\bx,L_0)$ by the random phase screen at $z=L_0$ and by propagating freely to $z=0$. The resulting expression is
\bea \label{UL}
U_{\omega_0}^{(1)}(\bx,0) &=& e^{ikL_0} \int_{\Rm^2} e^{i k |\bx-\by|^2/2L_0} e^{i k S_0(\by)}U_{\omega_0}^{(1)}(\by,L_0) P(\by)d\by,
\eea
where $S_0(\by)=\sigma_0 V_0(\by)$ is also a diffuser optical path length, for $V_0$ a mean-zero real-valued random field with unit variance. For an illumination with label $\ell$, the field $H_{\omega_0}^{(\ell)}(\bx,0)$ used in the reflection matrix is then directly given by $H_{\omega_0}^{(\ell)}(\bx,0)=(U_{\omega_0}^{(1)}(\bx,0))^2 e^{-4i k f}$.

The SHG field $\calG_{2\omega_0}(\bx,2z_s;\bx_j,0)$ in backpropagation is obtained in a similar manner, and we find
\bea \label{bG1}
\calG_{2\omega_0}(\bx,2z_s;\bx_j,0)&=&e^{i2k(2f-L)} e^{i \frac{2k L}{2 f^2} |\bx|^2} \int_{\Rm^2} e^{-\frac{i 2k }{f}\bx \cdot \by} e^{i2k S_1(\by)} \calG_{2\omega_0}(\by,L;\bx_j,0)  P(\by) d\by\\
\label{bG2} \calG_{2\omega_0}(\bx,L;\bx_j,0) 
&=&e^{i2k L} \int_{\Rm^2} e^{i \frac{2k }{2 L_1} |\by-\bx|^2}  e^{i \frac{2k }{2 L_0} |\by-\bx_j|^2} e^{i2kS_0(\by)}  P(\by) d\by.
\eea
It follows that $G_{2\omega_0}(\bu;\bx_j)=\calG_{2\omega_0}(\bu,2z_s;\bx_j,0) e^{-4ikf}$.

The speckle fields could be generated in a different manner. Instead of placing random phase screens, one could consider propagating the light through a random medium between $z=0$ and $z=L$. This would require the costly resolution of a full 3D high frequency problem, for somewhat little gain for us since the only characteristics of the speckle that matter in our imaging method are its spatial transverse correlation length and the memory length.

In the simulations, the function $P$ is the indicator function of a square of side $W=500 \lambda$ centered at zero, for $\lambda$ the central wavelength associated with the angular frequency $\omega_0$. Regarding the computation of $U_{\omega}^{(1)}(\bx,L)$, the domain of integration in $\by$ is chosen so that the variable $\bx$ (here the dual Fourier variable of $\bu=\by/\lambda f$) belongs to the square of side $W$. If the latter is discretized with a stepsize $h$, then the Fourier transform is calculated over a square of length $1/h$ centered at zero in the variable $\bu$. The function $P_{\rm{SLM}}(\by)$ is the indicator function of a disk of radius $\mathrm{NA}  f $, where $\mathrm{NA}$ is the NA of the microscope. The resolution in absence of random screens is then the usual diffraction-limited resolution $\lambda/ 2 \mathrm{NA}$. We set $\mathrm{NA}=0.75$.

\paragraph{Characteristic lengths.} The field on the SLM is of the form $\exp(i \sigma_{\rm{SLM}} V_{\rm{SLM}})$, where $V_{\rm{SLM}}$ is a mean-zero real-valued random field with unit variance. The random field $V_{\rm{SLM}}$ (resp. $V_0$ and $V_1$) is defined by a random Fourier series: we take the real part of the inverse Discrete Fourier Transform  
of independent random complex coefficients supported on a disk of radius of maximal wavenumber $\ell_{\rm{SML}}^{-1}$ (resp. $\ell_0^{-1}$ and $\ell_1^{-1}$). The lengths $\ell_{\rm{SLM}}$, $\ell_0$ and $\ell_1$ are the correlation lengths of the associated random fields, and the shortest wavelengths included the fields are then $2 \pi \ell_{\rm{SLM}}$, $2\pi \ell_0$ and $2\pi \ell_1$. The random Fourier coefficients are of the form $r e^{i \theta}$, where $r$ and $\theta$ are drawn uniformly in $[-1/2,1/2]$ and $[0, 2\pi]$. The constructed random fields are stationary, in the sense that $V_{\rm{SLM}}(\bx)$ and $V_{\rm{SLM}}(\bx+\by)$ e.g. have the same distributions for all $\bx,\by \in \Rm^2$. Note as well that when the discretization is sufficiently fine and $\ell_0^{-1}$, $\ell_1^{-1}$, $\ell_{\rm{SLM}}^{-1}$ are sufficiently large, the central limit theorem implies that $V_0$, $V_1$ and $V_{\rm{SLM}}$ are approximately Gaussian fields. This is the case in our simulations where we set $h=W/2^{11}=W/2048$ and $\ell_0=\ell_1=\ell_{\rm{SLM}}=2\lambda$. We will then treat the fields as Gaussian in our calculations in Appendices \ref{app:memory}, \ref{app:corr}, \ref{app:linde}. In more detail, we have for instance for $V_0$ (the expression below is obtained in the continuous limit of the Fourier series and is therefore only an accurate approximation),
$$
\E\{V_0(\bx)V_0(\by)\}=R_0(\bx-\by)=R_0(r)=\frac{2 J_1(\ell_0^{-1} r)}{\ell_0^{-1} r}, \qquad r=|\bx-\by|,
$$
where $J_1$ is the first order Bessel function of the first kind. Similar expressions hold for $V_1$ and $V_{\rm{SLM}}$ by adjusting the correlation length. Using the Gaussian property, we find for the random screen
$$
\E\{e^{i k \sigma_0 (V_0(\bx)-V_0(\by))}\}=e^{-k^2 \sigma_0^2(1-R_0(\bx-\by))}.
$$
In a speckle regime, we expect fluctuations to be quite strong, and therefore $e^{-(k \sigma_0)^2} \ll 1$. The correlation length of the phase screen is then found by expanding $R_0$ around the origin as (we use the fact $R_0$ is even),
$$
1-R_0(|\bx-\by|)=-\frac{1}{2}R_0''(0)|\bx-\by|^2+o(|\bx-\by|^2).
$$
A short calculation shows that $
R_0''(0)=-1/4\ell_0^2$, and, as a consequence 
$$
e^{-k^2\sigma_0^2(1-R(\bx-\by))} \simeq e^{-k^2 \sigma_0^2R_0''(0)|\bx-\by|^2/2}=e^{-|\bx-\by|^2/l_{c,0}^2}
$$
where
$$l_{c,0}=\frac{\sqrt{2}}{k\sigma_0 \sqrt{-R''(0)}}=\frac{2\sqrt{2} \ell_0}{k \sigma_0}.$$
We define $l_{c,1}$ and $l_{c,\rm{SLM}}$ analogously. When $e^{-(k \sigma_0)^2} \ll 1$, the phase screens then decorrelate fast at the scales $l_{c,0}$ and $l_{c,1}$, which is a property that will be used when establishing further the Gaussian nature of the fields $U_{\omega_0}^{(1)}(\bx,0)$ and $\calG_{2\omega_0}(\bx,2z_s;\bx_j,0)$.

The wavenumber $k$ is doubled for the SHG field, and we introduce $\ell_{c,0}=l_{c,0}/2$ as well as $\ell_{c,1}$ and $\ell_{c,\rm{SLM}}$ similarly. We set $k\sigma_0=k\sigma_1=\sigma_{\rm{SLM}}=2$ in our simulations, and with $\ell_0=\ell_1=\ell_{\rm{SLM}}=2\lambda$, we find $l_{c,0}=l_{c,1}=l_{\rm{SLM}}=2\sqrt{2} \lambda$. The averages of $U_{\omega_0}^{(1)}(\bx,0)$ and $\calG_{2\omega_0}(\bx,2z_s;\bx_j,0)$ involve
$$
\E\{e^{i k \sigma_0 V_0(\bx)}e^{i k \sigma_1 V_1(\by)}\}=e^{-\frac{k^2}{2} (\sigma_0^2+\sigma_1^2)}=e^{-4}\simeq 0.02,
  $$
  so from a practical viewpoint, $U_{\omega_0}^{(1)}(\bx,0)$ and $\calG_{2\omega_0}(\bx,2z_s;\bx_j,0)$ are essentially mean zero.

We show in Appendix \ref{app:corr} that the speckle correlation lengths on the way forward of the fundamental field is
$$
\ell_{s,\rm{in}}=\min\left( \lambda/2 \mathrm{NA},l_{c,0},L_0l_{c,1}/L\right)
$$
while that of the SHG field is
$$
\ell_{s,\rm{out}}=\min\left( \lambda f/2D,\ell_{c,0},L_0\ell_{c,1}/L\right).
$$
Above, we assumed measurements on the camera are performed over a square of side $D$ centered at zero. We set $D=402 h \simeq 100 \lambda$ and $L_0=0.8f=400 \lambda$ and $L_1=0.1f=50\lambda$. With such choices, we find
$$
\ell_{s,\rm{in}}=\lambda/2 \mathrm{NA} \simeq 0.66 \lambda, \qquad \ell_{s,\rm{out}}=L_0\ell_{c,1}/L \simeq 1.25 \lambda.
$$
Regarding the memory effect, we derive an analytical expression in Appendix \ref{app:memory}, which specialized to our numerical value yields

$$\ell_{\rm{me}}=\frac{\sqrt{2} \ell_1 L_0}{k \sigma_1 L_1} \simeq 11.31 \lambda.$$

Table \ref{tab} gathers the values of the different parameters used in the simulations.
\begin{table}[h!]
\begin{center}
\begin{tabular}{|c|c|c|c|c|c|c|c|c|c|c|c|c|c|}
    \hline
     $W$ & $h$ &$z_s$ & $L_0$& $L_1$ & $D$ & $\ell_0$& $\ell_1$ & $\ell_{\rm{SLM}}$ & $k \sigma_0$ & $k \sigma_1$ & $\sigma_{{\rm SLM}}$ &$\mathrm{NA}$ & $\tau$ \\
     \hline
     500$\lambda$ & $W/2048$ &500$\lambda$ & 0.8$z_s$& 0.1$z_s$ & $402 h$& $2\lambda$ & $2\lambda$ & $2\lambda$ & 2 & 2 & 2&0.75& 1\\
     \hline
\end{tabular}
\end{center}
\caption{Parameters used in the simulations.}
\label{tab}
\end{table}

\paragraph{Speckle regime.} The fact that $U_{\omega_0}^{(1)}(\bx,0)$ and $\calG_{2\omega_0}(\bx,2z_s;\bx_j,0)$ are in a speckle regime is a consequence of the central limit theorem (CLT) as we sketch informally below. Consider first integrals of the form
$$
I(\bx)=l_N^{-2}\int_{|\by|\leq W/2}e^{-\frac{i 2 \pi }{\lambda d_0}\bx \cdot \by} N(\by/l_N) d\by,
$$
where $N$ is a complex-valued random field with correlation length equal to one, and where $d_0,l_N$ are some distances. For some number $\mu$ and $e_1=(1,0)$, set $\bx= e_1 \mu \lambda d_0/l_N $ and assume for simplicity that $W/2l_N$ is an integer that we denote $M$. The square $|\by|\leq W/2l_N$ can then be partitioned into $4M^2$ squares $S_1$ of side 1. Denoting by $\by_\ell$ the centers of these squares, we have, for $\by=(y_1,y_2)$,
\bee
I(\bx)&=&\int_{|\by|\leq W/2l_N}e^{-i 2 \pi \mu y_1 }  N(\by) d\by=\sum_{\ell=1}^{4 M^2}\int_{S_1}e^{-i 2 \pi \mu (\by_\ell+\by)_1}  N(\by_\ell+\by) d\by\\
&=&\sum_{\ell=1}^{4 M^2} X_\ell(\mu).
\eee
Note that $X_\ell(\mu) \sim 0$ when $2 \pi |\mu| \gg 1$, so we only consider $\mu$ such that $2 \pi |\mu|\leq 1 $. Here, $I(\bx)$ is a placeholder for $\calG_{2\omega_0}(\bx,2f;\bzero,0)$ and $U_{\omega_0}^{(1)}(\bx,0)$, and it is clearly observed in the simulations that these yields are small when $|\bx|$ is far from the origin (this is quantified by the distance $\lambda d_0/l_N$ for appropriate $d_0$ and $l_N$). The random variables $X_\ell(\mu)$ are neither independent nor identically distributed, so the standard form of CLT cannot be used. Under the Lindeberg condition, and assuming the $X_\ell(\mu)$ are sufficiently weakly correlated, the CLT holds when $M \gg 1$, see e.g. \cite{CLTUtev}. The integral $I(\bx)$ has then approximately a complex Gaussian statistics, and it is circularly symmetric when $\E\{I\}=\E\{I^2\}=0$.

The above result can be applied to show that $\calG_{2\omega_0}(\bx,2f;\bzero,0)$ and $U_{\omega_0}^{(1)}(\bx,0)$ have circular complex Gaussian statistics by setting respectively $d_0=f/2$ and $N(\by/l_N)=e^{i2kS_1(\by)}  P(\by)\calG_{2\omega_0}(\by,L;\bzero,0)$, and $d_0=L_0$, $N(\by/l_N) =e^{i k |\by|^2/2L_0} e^{i k S_0(\by)} U_{\omega_0}^{(1)}(\by,L_0) P(\by)$. We only comment on the first case, where the terms $e^{i2kS_1(\by)}$ and $\calG_{2\omega_0}(\by,L;\bzero,0)$ are statistically independent. The phase screen $e^{i2kS_1(\by)}$ decorrelates fast when $e^{-k^2 \sigma_1^2} \ll 1$, as discussed above, and at the scale $\ell_{c,1}$. The latter is much smaller than the window size $W$, and as a consequence indeed $M \gg 1$. We set $l_N=\ell_{c,1}$, and the $X_\ell(\mu)$ are sufficiently mixing to apply the CLT. Finally, we address the non-identical character of the $X_\ell(\mu)$, which  stems from the non-stationarity of $e^{-i 2 \pi \mu (\by_\ell)_1} \calG_{2\omega_0}(\bx,L;\bzero,0)$ (the term $e^{i2kS_1(\by)}$ is stationary). Non-stationarity arises from deterministic purely oscillatory terms, and we show in Appendix \ref{app:linde}  that the Lindeberg condition holds for $X_\ell(\mu)$. The CLT then establishes the Gaussian property. Circularity is a consequence of the fact that moments of the random screens of the form  $\E\{e^{ik(S_j(\by)+S_j(\by'))}\}$ are small since $e^{-k^2 \sigma_j^2} \ll 1$, $j=0,1$.

\begin{figure}[h!]
\begin{center}
  \includegraphics[scale=0.28]{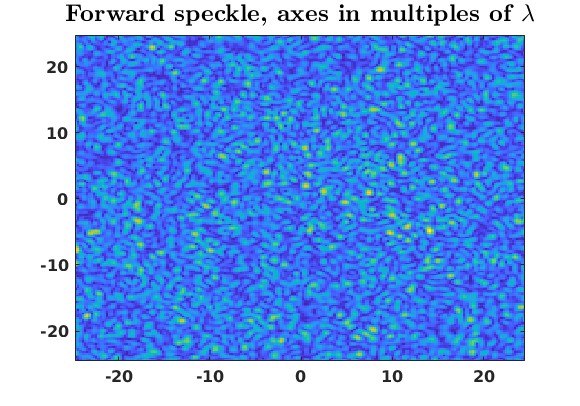}
  \includegraphics[scale=0.27]{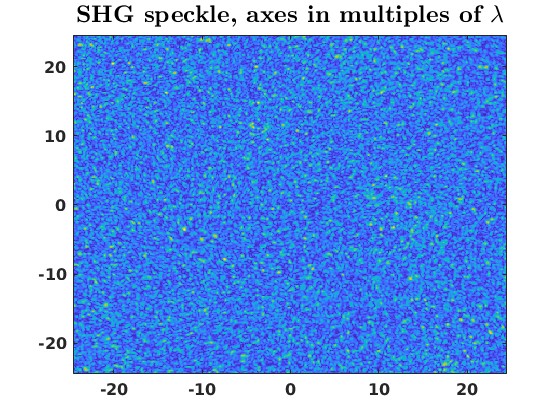} 
\end{center}
\caption{Forward (left) and SHG speckle (right). The SHG speckle has finer grains since its wavelength is half that of the forward field.}
\label{fig:speck}
\end{figure}

We represent in Figure \ref{fig:speck} the forward and SHG speckles obtained in our simulations. The forward speckle is represented on the sample plane and the SHG one on the camera. Note the finer speckle grains for the SHG field due to its wavelength being half that of the forward field.


\subsection{Resolution analysis} \label{reso}
We investigate in this section the resolution limit in ideal cases. We first establish the resolution limit introduced by the deconvolution. From a numerical viewpoint, an important remark is that the measured field at $z=2f$ lives on a different grid than the field on the sample plane at $z=0$. This means that the resolution is limited by the pixel size on the camera, and we show numerically that this limit is achieved in our setting. The pixel size on the sample plane is $h \simeq 0.25 \lambda$ with $N_{{\rm grid}}=2048$ points in each direction, while the one on the camera is $(\lambda N_{\rm{grid}}/ 2 W) h=2.048 h$. (This is established by inspecting Fourier variables in the discrete Fourier transform). If two scatterers are located at $\bx_i$ and $\bx_j$ in the sample plane, we show in Appendix \ref{app:memory} that the deconvolution reconstructs the shift $\alpha_{\rm{me}} (\bx_i -\bx_j)$ with $\alpha_{\rm{me}}=f/L_0=1.25$ on the camera grid. This shift then needs to be properly rescaled to capture the original separation in the sample plane. Suppose for instance $\bx_i$ and $\bx_j$ are aligned on the $x$ axis. If the deconvolution estimates a shift of $n$ pixels on the camera for a distance of about $2hn$, the distance between the scatterers on the sample plane is about $2hn/ \alpha_{\rm{me}}$. To summarize, distances estimated on the camera need (theoretically) to be divided by $\alpha_{\rm{me}}=1.25$. The best numerical factor we find is $1.2$, which is about $4\%$ off the theoretical value. We will systematically use the numerical value for representing the reconstructions, leading to a pixel size on the reconstructions of $2.048h/1.2 \simeq 0.42 \lambda$.

\begin{figure}[h!]
\begin{center} 
  \includegraphics[scale=0.135]{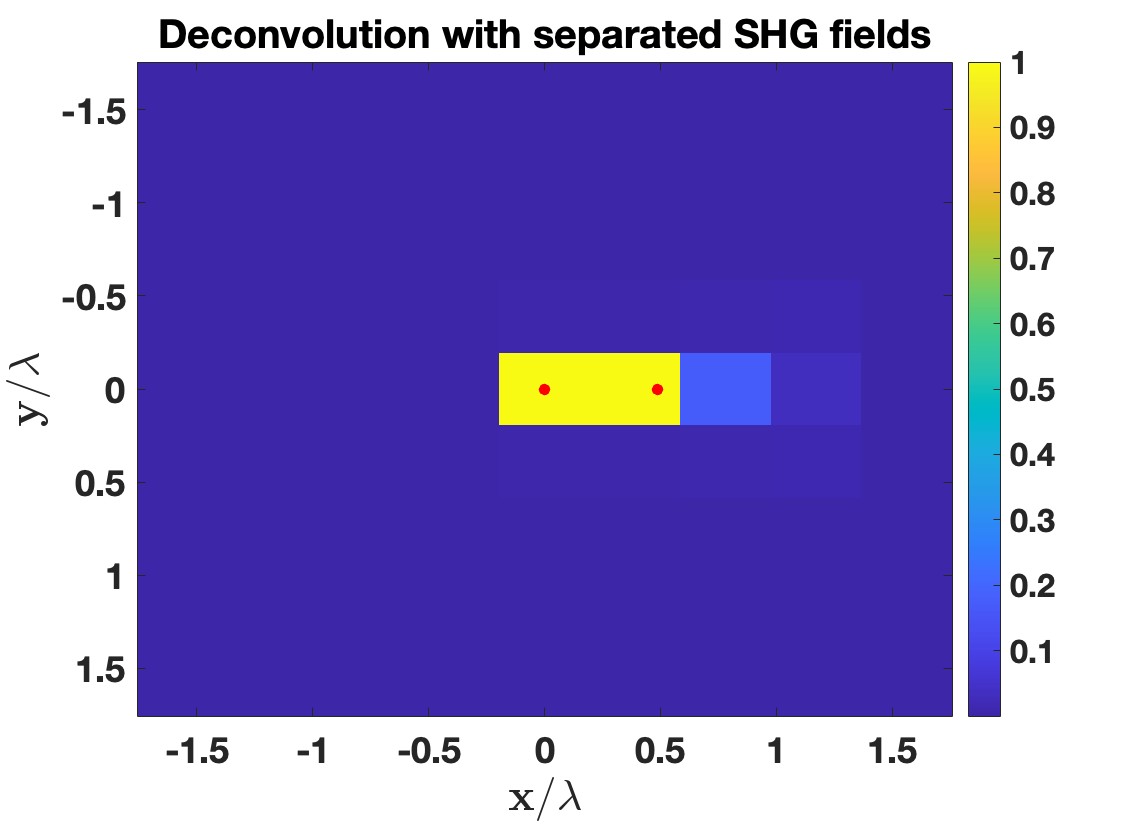}\hspace{-0.43cm}
  \includegraphics[scale=0.135]{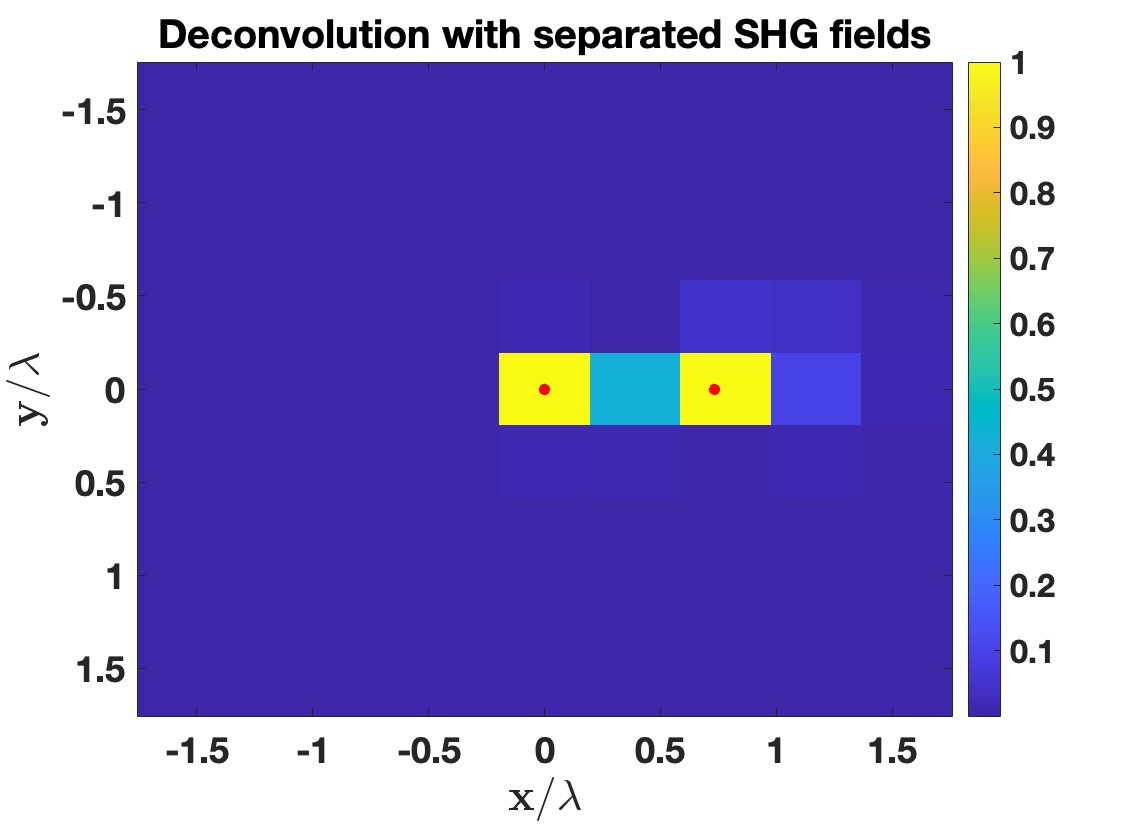}\hspace{-0.43cm}
    \includegraphics[scale=0.135]{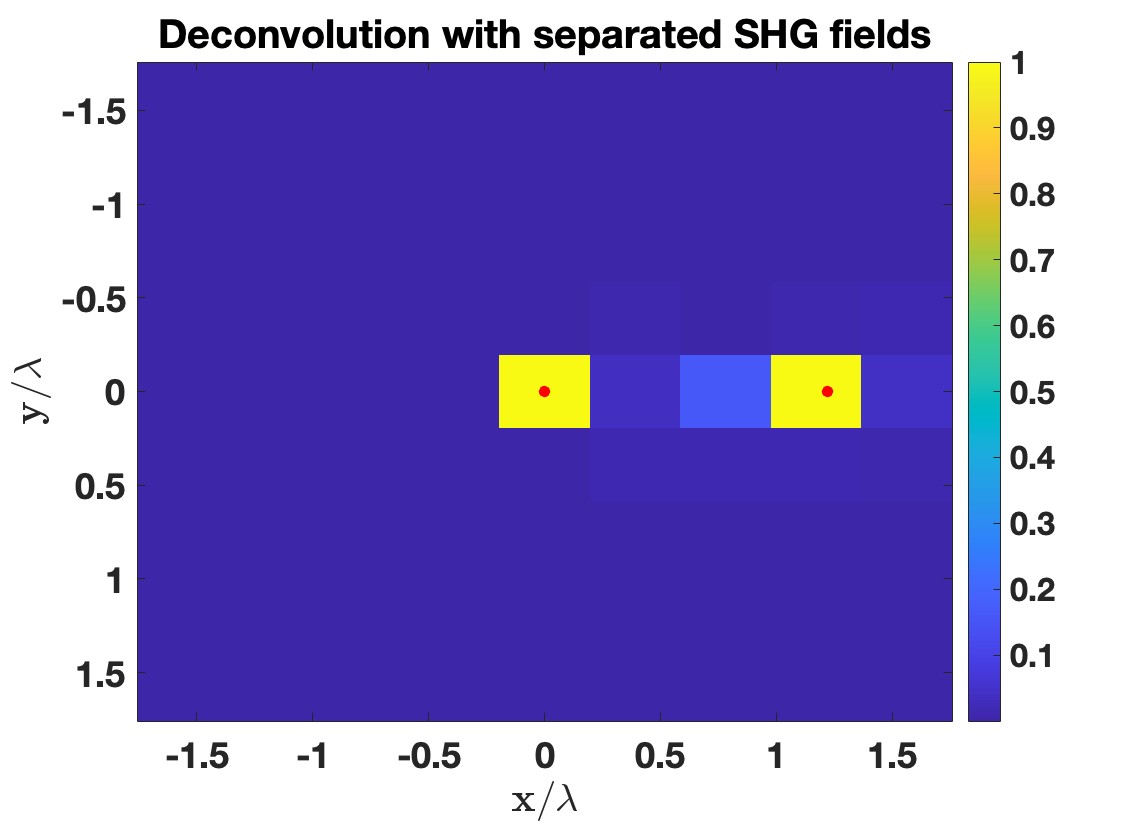} 
  \end{center}
  \caption{Deconvolution with perfectly separated SHG speckles. Left: $d=2h$. Center $d=3h$. Right $d=5h$. The red dots show the exact locations of the scatterers. The deconvolution achieves the resolution of the camera pixel size.}
  \label{fig:deconlimit}
\end{figure}

In order to test the deconvolution, we assume we are given two perfectly separated SHG fields $G_{2 \omega_0}(\bx;\bx_i)$ and $G_{2 \omega_0}(\bx;\bzero)$ emanating from points scatterers at $\bx_i$ and $\bzero$ with same susceptibilities. We set $\bx_i$ on the grid and on the $x$ axis with $|\bx_i|=d$. Since the pixel size on the camera is about twice that on the sample plane (i.e. $h$), the best we can achieve in our setting is locating scatterers that are at least $3h$ away from each other. This is precisely what is shown in Figure \ref{fig:deconlimit}. We have $\tau=1$ in all of our reconstructions. The deconvolution is robust with respect to this choice, and images do not qualitatively change when increasing or decreasing $\tau$ moderately. 

We clearly observe on Figure \ref{fig:deconlimit} that the scatterers are not resolved for $d=2h$, but are when $d\geq 3h$. The conclusion is hence that the deconvolution achieves the resolution of the camera pixel size.

We consider now a similar  situation with two scatterers in a real imaging context to test the separation limit of the ICA. Since the decorrelation length $\ell_{s,\rm{in}}$ of the input field is between $2h$ and $3h$, we expect the algorithm to be able to separate fields emanating from scatterers at least $3h$ apart. This is what we observe in our simulations, provided the number of illuminations is sufficiently large, here $N_r=30$, and the center panel in Figure \ref{fig:deconlimit} is reproduced in the case of non-separated fields.

\begin{figure}[h!]
\begin{center}
  \includegraphics[scale=0.135]{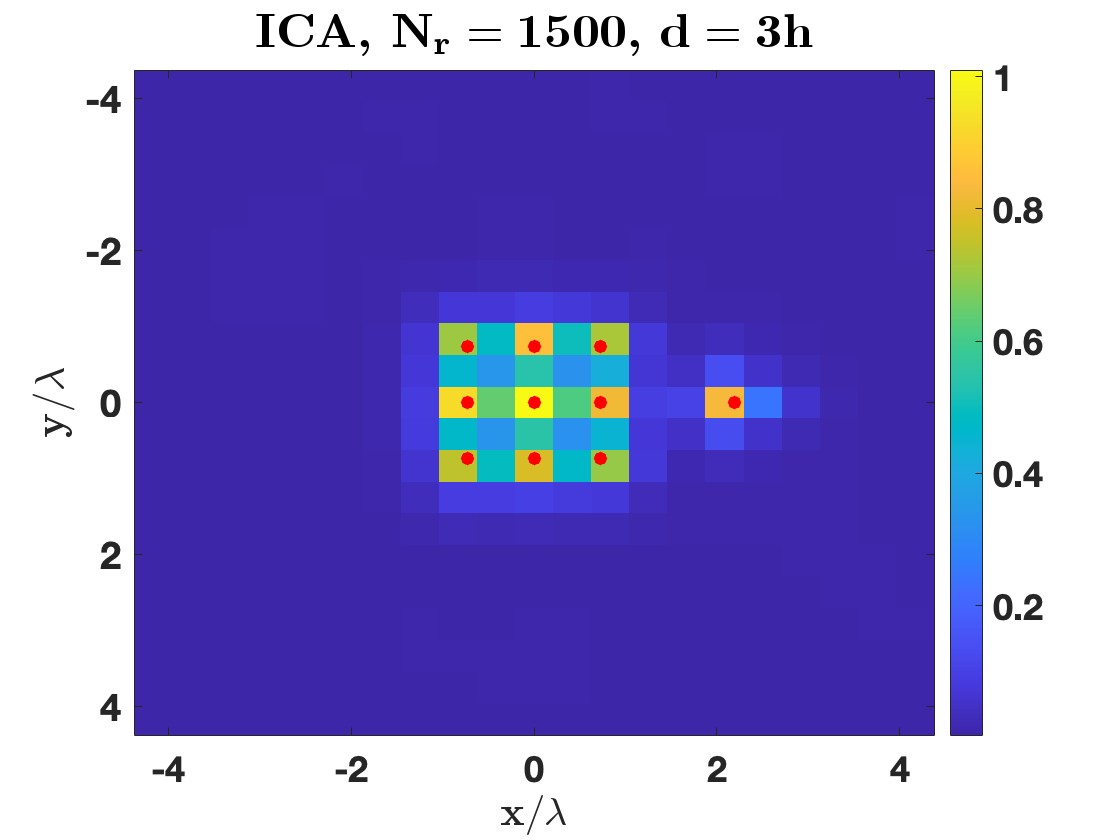}\hspace{-0.43cm}
  \includegraphics[scale=0.135]{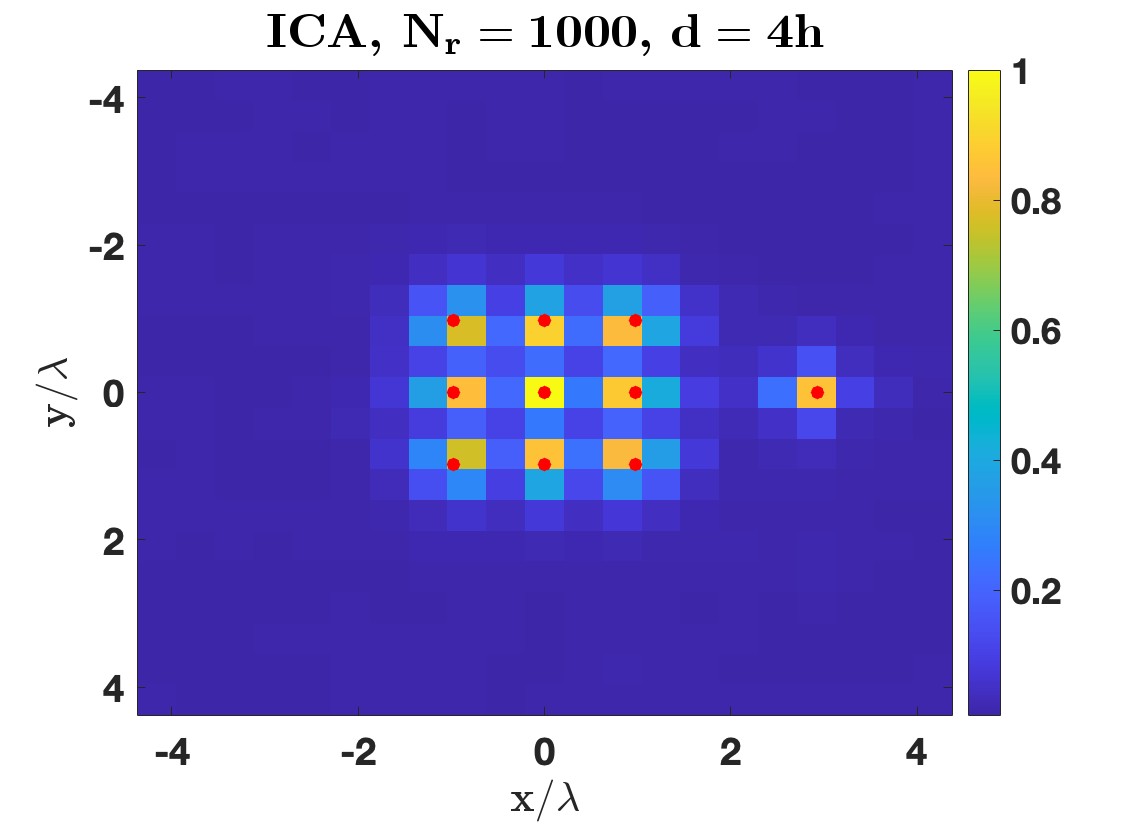}\hspace{-0.43cm}
  \includegraphics[scale=0.135]{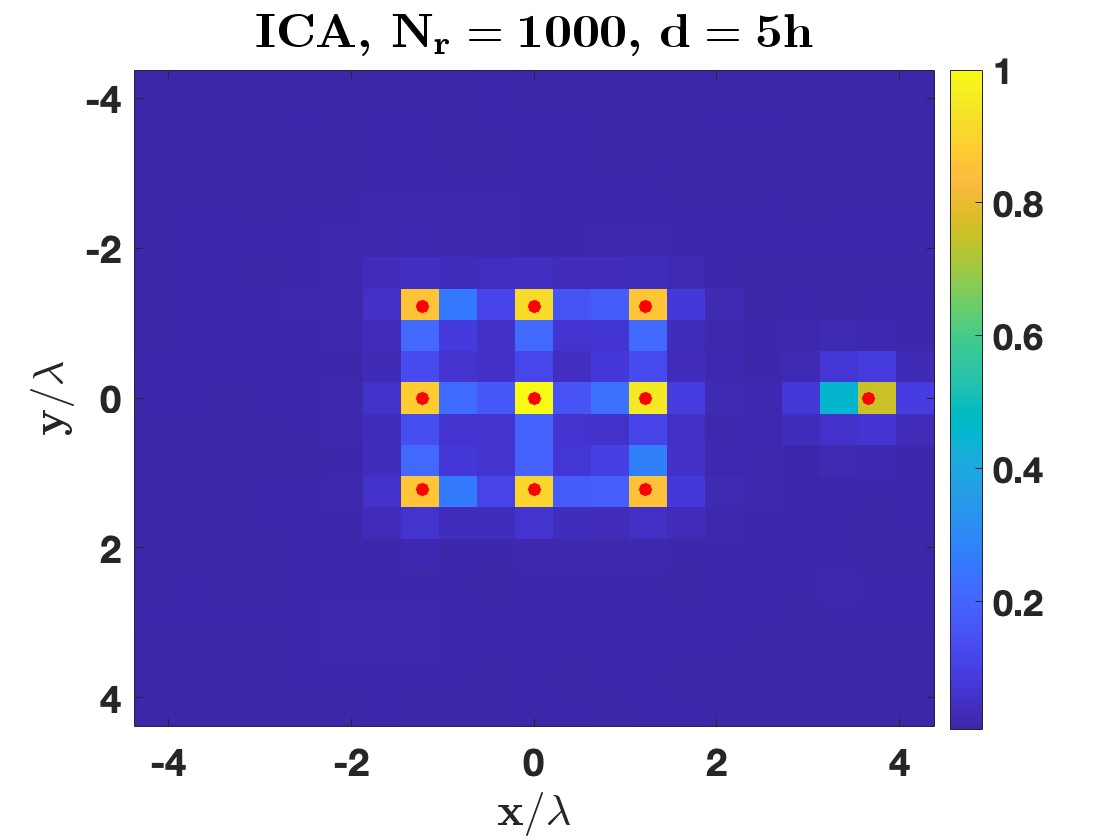}
  \caption{Reconstructions with 10 scatterers. Left: minimal distance $d$ is $3h$ with $N_r=1500$. Center: $d=4h$, $N_r=1000$. Right: $d=5h$, $N_r=1000$. The red dots show the exact locations of the scatterers.}
  \label{fig:10scat}
\end{center}
\end{figure}

In this simple setting with two scatterers, our method is thus able to image scatterers that are at least $\ell_{s,\rm{in}}$ apart, as expected. When there are a larger number of closely spaced scatterers, the number of illuminations needs to be increased to obtain a good separation. In Figure \ref{fig:10scat}, we image 10 identical scatterers, 9 of which are close together. When the minimal distance $d$ between the scatterers is $3h$, we need at least $N_r=1500$ to find the correct locations with a low quality image. When $d=4h$ and $d=5h$, $N_r=1000$ is sufficient and the scatterers are very well imaged.  The general trend is that fewer illuminations are needed the farther apart the scatterers are.

\begin{figure}[h]
\begin{center}
  \includegraphics[scale=0.135]{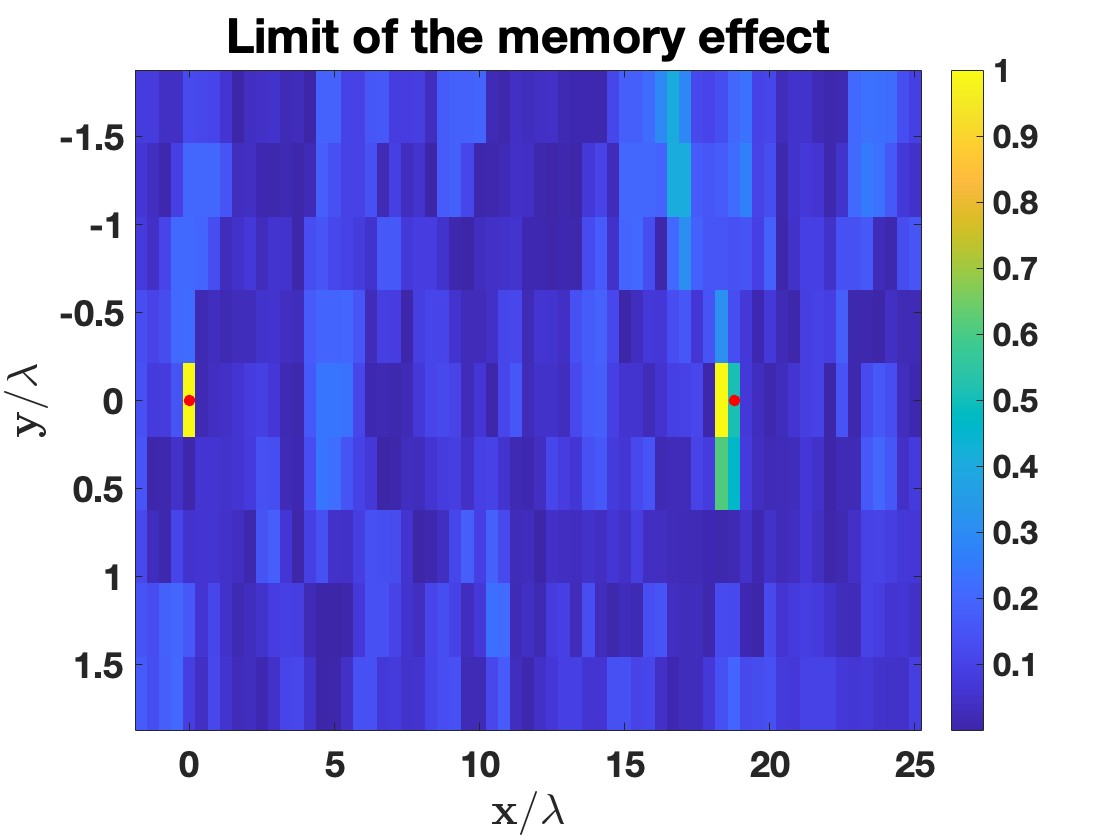}
\end{center}
\caption{Numerical limit of the memory effect. Beyond about $18 \lambda \simeq 1.6 \ell_{\rm{me}}$, background noise increases and the scatterers may not be well located.}
\label{fig:limM}
\end{figure}

To conclude this section, we investigate the limit of the  memory effect: the theoretical memory length $\ell_{\rm{me}}$ quantifies the correlation length between $|G_{2\omega_0}(\bu;\bx_j)|$ and $|G_{2\omega_0}(\bu+\alpha_{\rm{me}} (\bx_i-\bx_j);\bx_i)|$, so that when $|\bx_i-\bx_j|=\ell_{\rm{me}}$, the correlation has changed by say half of its value. In the simulations, the deconvolution is still able to estimate the shift past $\ell_{\rm{me}}$, and we estimate here how much farther. Figure \ref{fig:limM} shows a limit of about $18 \lambda \simeq 1.6 \ell_{\rm{me}}$, where background noise increases and the scatterer is wrongly located by one pixel. As can be expected, we have observed that the farther the scatterers, the larger the error on the estimated shifts.

\subsection{Imaging with the SVD only} \label{imSVD}

We briefly investigate in this section reconstructions without the ICA using the SVD only. We have discussed in Section \ref{sec:SVD} the fact that the SVD alone separates poorly nearly orthogonal fields emanating from scatterers with similar susceptibilities, and show here results supporting the analysis. 

When the scatterers are located at distances larger than the correlation lengths of the incoming and SHG speckle fields, the latter enjoy some orthogonality. When for instance $|\bx_i-\bx_j| \gg \ell_{s,\rm{out}}$, we have indeed, according to the law of large numbers, and using the fact that the fields oscillate fast compared to the overall domain of integration $S_D$ (the square of side $D$ centered at zero),
\bee
\frac{1}{D^2}\int_{S_D} G_{2 \omega_0}(\bx;\bx_i)G^*_{2 \omega_0}(\bx;\bx_j) d\bx &\simeq& \E\{G_{2 \omega_0}(\cdot;\bx_i)G^*_{2 \omega_0}(\cdot;\bx_j)\}\\
&\simeq& \E\{G_{2 \omega_0}(\cdot;\bx_i)\}\E\{G^*_{2 \omega_0}(\cdot;\bx_j)\} \simeq 0.
\eee
A similar relation holds for the fundamental field. 

\begin{figure}[h!]
\begin{center}
  \includegraphics[scale=0.27]{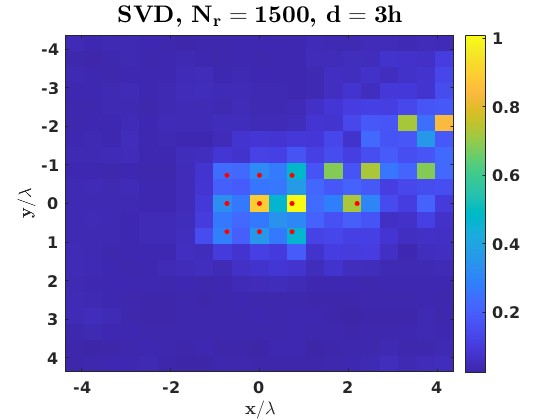}\hspace{-0.43cm}
  \includegraphics[scale=0.27]{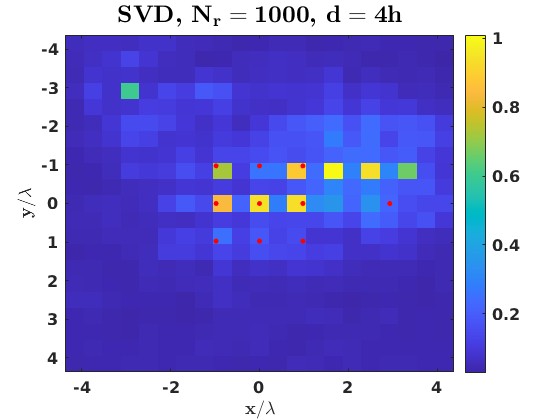}\hspace{-0.43cm}
  \includegraphics[scale=0.27]{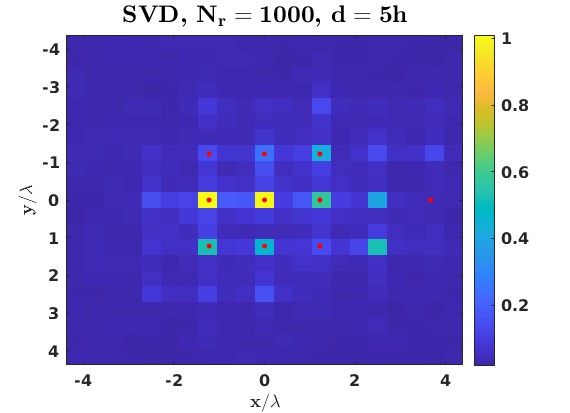}
  \caption{Reconstructions for identical scatterers with SVD only without ICA separation. The red dots show the exact locations of the scatterers. The SVD alone is not able to separate the fields well enough for a well resolved image.}
  \label{fig:10scatsvd}
\end{center}
\end{figure}

\begin{figure}[h!]
  \begin{center}
    \includegraphics[scale=0.139]{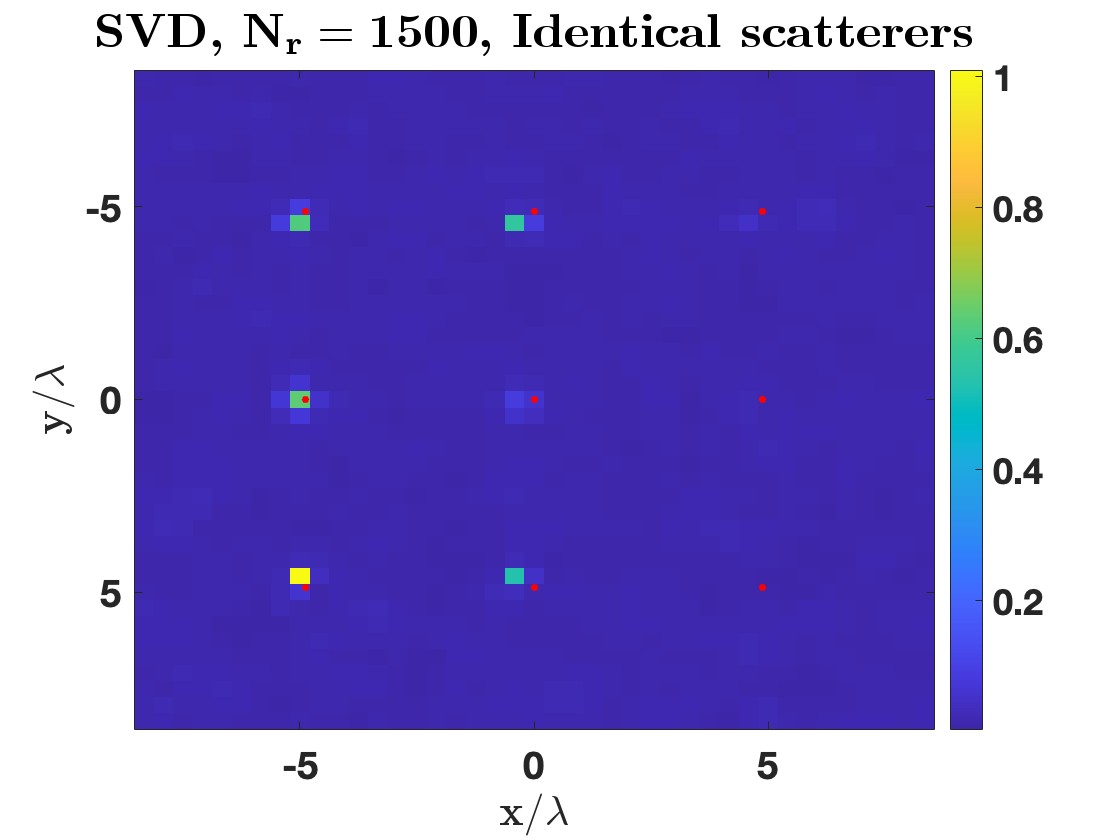}
  \includegraphics[scale=0.135]{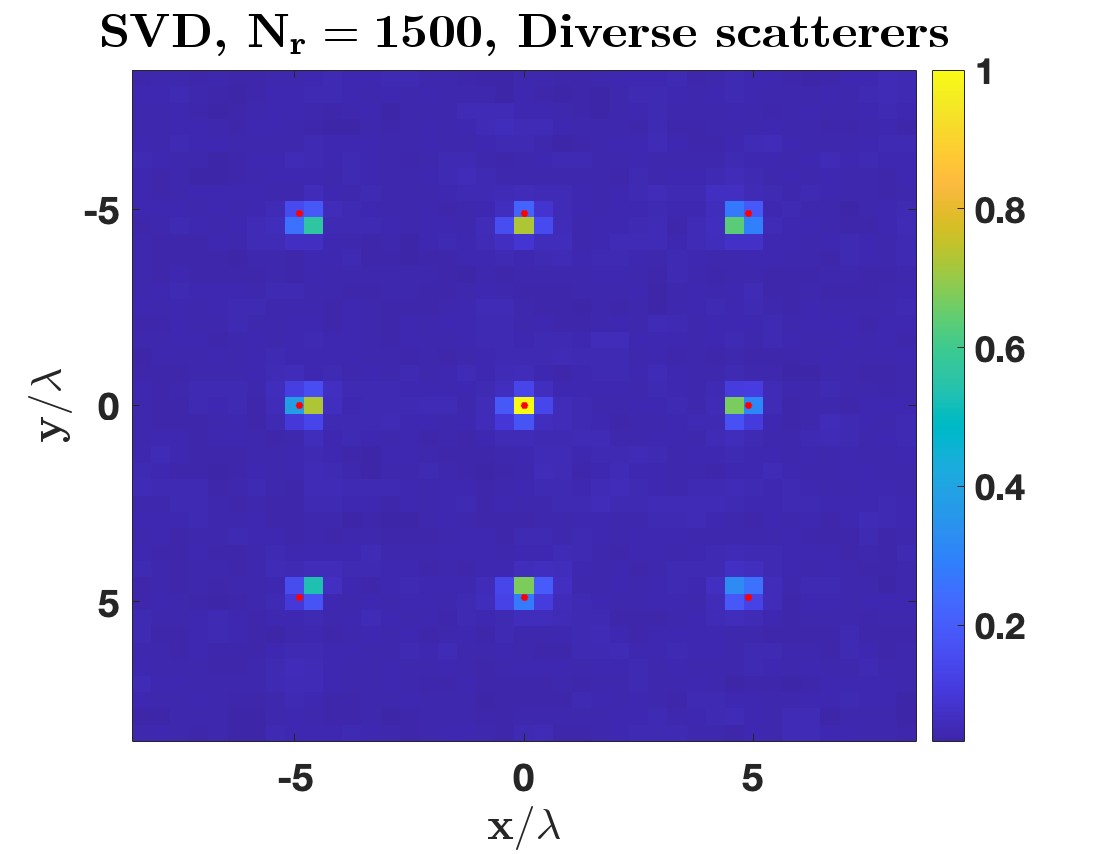}
     \end{center}
  \caption{Reconstructions for identical scatterers with SVD only without ICA separation. Left: identical scatterers with minimal separation $d=20h=5 \lambda$. Right: diverse scatterers with random susceptibilities for $d=20h$. The scatterers are well resolved in the latter case since the speckle bond resonance effect is mitigated.}
  \label{fig:10scatSVD2}
\end{figure}

In the same setting as Figure \ref{fig:10scat}, we show in Figure \ref{fig:10scatsvd} the obtained images, which are clearly not resolved. We have to increase the distance to $d=20h$ to locate some scatterers, albeit with a mediocre image. As expected, the image is significantly improved when the susceptibilities of the scatterers are sufficiently diverse (here chosen at random as $r e^{i\theta}$, $r$ uniform in $[1/2,3/2]$ and $\theta$ uniform in $[0,2\pi]$), to overcome the speckle bond resonance; see Figure \ref{fig:10scatSVD2}.


\subsection{Reconstructions of discrete scatterers} \label{scat}
\begin{figure}[h!]
\begin{center}
  \includegraphics[scale=0.27]{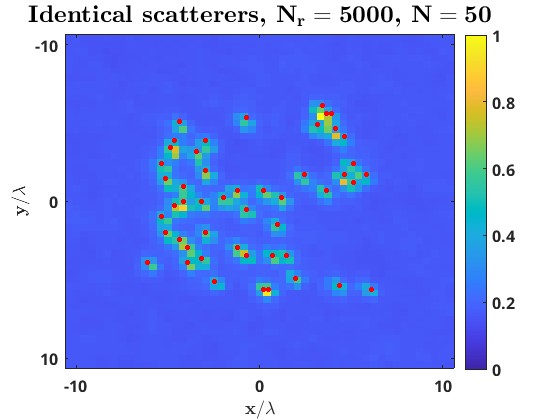}\hspace{-0.43cm}
  \includegraphics[scale=0.27]{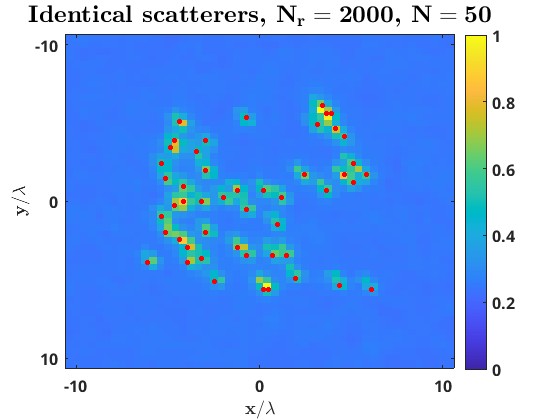}\hspace{-0.43cm}
  \includegraphics[scale=0.27]{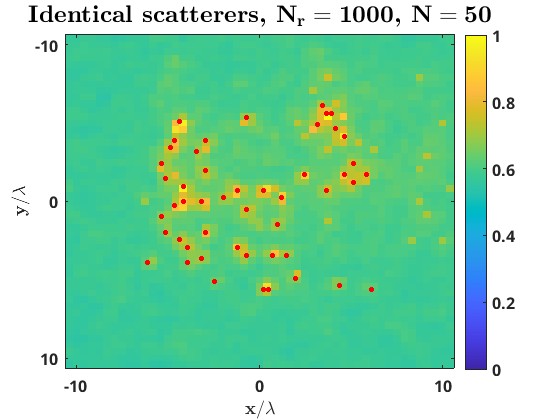}
  \caption{Imaging of 50 identical scatterers with random locations. Left: $N_r=5000$ illuminations. Center $N_r=2000$. Right: $N_r=1000$. The red dots show the exact locations of the scatterers. The first two cases exhibit comparable results, while the last one yields a significantly lower signal-to-noise ratio. The denoising parameter $\sigma$ is set to zero.}
  \label{fig:50scat}
\end{center}
\end{figure}

This section is dedicated to the imaging of randomly placed scatterers. We consider a first case with $N=50$ identical scatterers drawn uniformly of a square grid of size $50h\times 50h \simeq 12.5\lambda \times 12.5 \lambda$. In Figure \ref{fig:50scat}, we vary the number of illuminations $N_r$, and keep the same color code to show the signal-to-noise ratio improvement as $N_r$ increases. The denoising parameter $\sigma$ in formula \ref{expI} is set to zero in Figure \ref{fig:50scat}.

The images show excellent reconstructions for $N_r \geq 2000$, which is about $40$ times more illuminations than scatterers. The relative distance between nearby scatterers tends to be very well estimated, while the relative distance between clusters of scatterers situated further apart can exhibit an error of a few pixels (we recall of size 0.42$\lambda$). This is particularly visible for the scatterers located at the periphery of the image, and the phenomenon is due in part to the way the images $I_i$ in \fref{expI} are merged: the distance estimated by the deconvolution between far apart scatterers is less accurate than that for close scatterers, and this distance is the quantity used for the merge, as explained in Section \ref{sec:TV}. So even though the relative positions of the scatterers within a cluster are accurately estimated, an error in the relative position of the whole cluster compared to a farther away reference scatterer induces an error on the entire cluster. 

\begin{figure}[h!]
\begin{center}
  \includegraphics[scale=0.27]{Figs/Id-50-5000-f0.jpg}\hspace{-0.43cm}
  \includegraphics[scale=0.27]{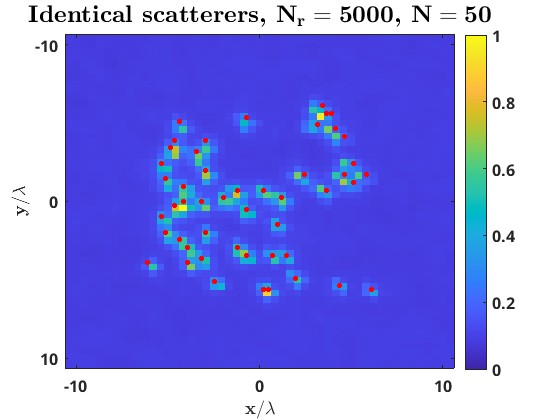}\hspace{-0.43cm}
  \includegraphics[scale=0.27]{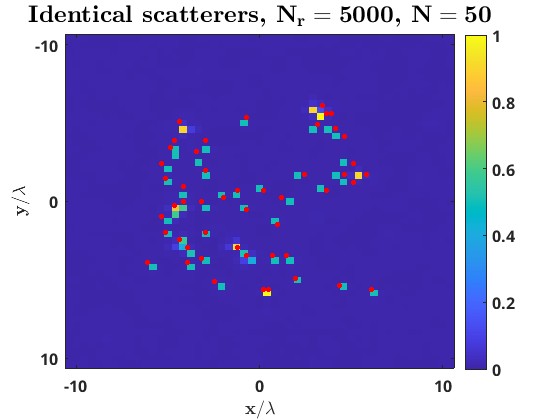}
   \caption{Imaging of 50 identical scatterers with random locations with $N_r=5000$ illuminations. Left: the denoising parameter $\sigma$ is set to zero. Center: $\sigma=0.4$. Right: $\sigma=0.8$. In the last case, the image consists of unique pixels giving the estimated scatterers locations.}
  \label{fig:50scat2}
\end{center}
\end{figure}

In Figure \ref{fig:50scat2}, we vary the denoising parameter $\sigma$. As $\sigma$ increases, so does the contrast and when $\sigma=0.8$ the image consists of (almost) unique pixels giving the estimated scatterers' locations.

Note that randomizinig the scatterers' susceptibilities barely changes the image as the ICA is immune to the speckle bond resonance.

\begin{figure}[h!]
\begin{center}
  \includegraphics[scale=0.27]{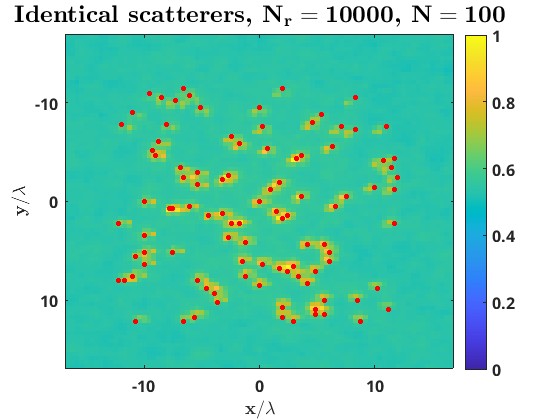}\hspace{-0.43cm}
  \includegraphics[scale=0.275]{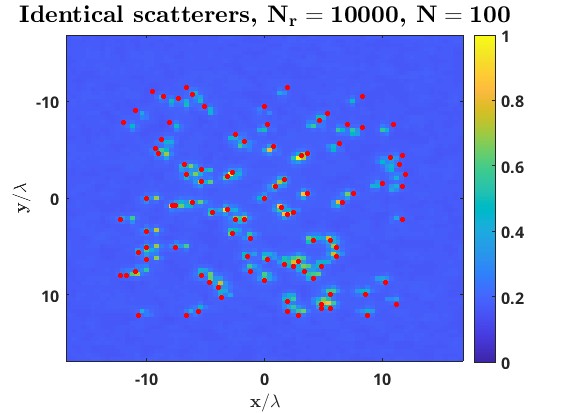}\hspace{-0.43cm}
  \includegraphics[scale=0.27]{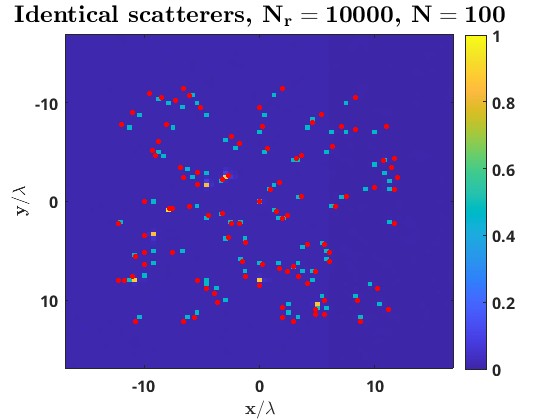}
  \caption{Imaging of 100 identical scatterers with random locations with $N_r=10000$ illuminations. Left: the denoising parameter $\sigma$ is set to zero. Center: $\sigma=0.4$. Right: $\sigma=0.8$. The red dots show the exact locations of the scatterers. }
  \label{fig:50scat3}
  \end{center}
\end{figure}

In Figure \ref{fig:50scat3}, we double the number of scatterers as well as the size of the window where the positions are drawn (now $25\lambda \times 25 \lambda$ and $N=100$). As in the case $N=50$, the reconstructions are excellent, with positions of clusters off possibly by a few pixels. 


\subsection{Reconstructions of continuous objects} \label{cont}
We image in this section continuous objects, meaning geometric forms built from scatterers placed at the grid size $h\simeq 0.25 \lambda$.

\begin{figure}[h!]
\begin{center}
  \includegraphics[scale=0.27]{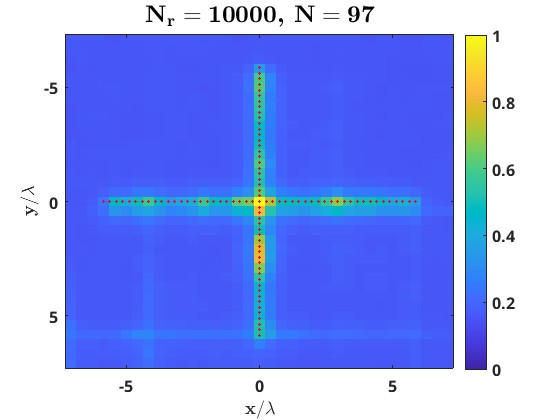}\hspace{-0.43cm}
  \includegraphics[scale=0.27]{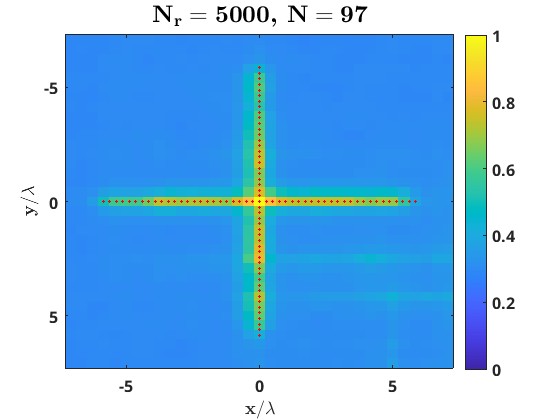}\hspace{-0.43cm}
  \includegraphics[scale=0.27]{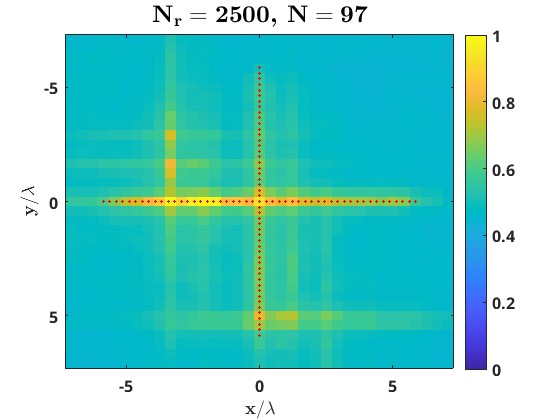}
  \caption{Imaging of a cross made with $N=97$ scatterers using all 97 singular vectors. Left: $N_r=10000$ realizations. Center: $N_r=5000$. Right: $N_r=2500$. The reconstructions improve with $N_r$.}
  \label{fig:crossN}
\end{center}
\end{figure}

The first object we consider is a cross made of 97 identical scatterers (24 in each branch plus the center). In Figure \ref{fig:crossN}, we keep all 97 singular vectors given by the SVD and vary the number of realizations. While the cross is reconstructed when $N_r=2500$, there a many artifacts and the quality of the image is poor. The latter improves as $N_r$ increases to $5000$ and $10000$.

\begin{figure}[h!]
\begin{center}
  \includegraphics[scale=0.27]{Figs/cross-5000-f0.jpg}\hspace{-0.43cm}
  \includegraphics[scale=0.27]{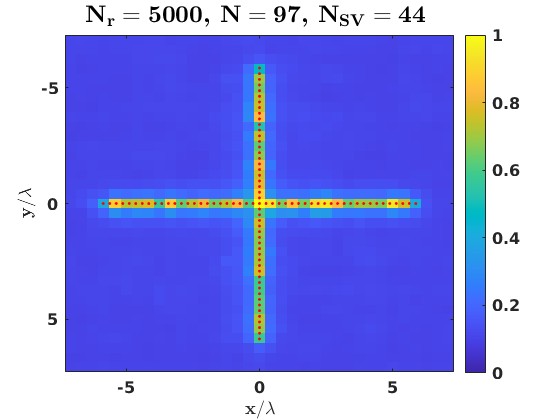}\hspace{-0.43cm}
  \includegraphics[scale=0.27]{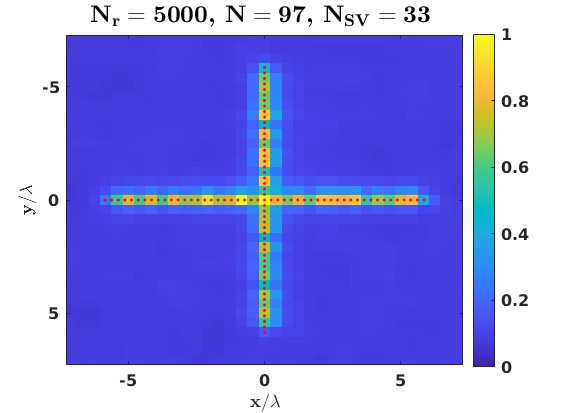}
  \caption{Imaging of a cross made with $N=97$ scatterers using only $N_{\rm{SV}}$ singular vectors for $N_r=5000$ realizations. Left: all singular vectors are used. Center: $N_{\rm{SV}}=44$ (those larger than 1$\%$ of the greatest singular value). Right: $N_{\rm{SV}}=33$ (those larger than 10$\%$ of the greatest singular value). The reconstructions improve when filtering out small singular values.}
  \label{fig:crossSV}
\end{center}
\end{figure}

In Figure \ref{fig:crossSV}, we vary the number of singular vector used for the reconstructions: when the scatterers are so close that the associated speckle fields are not orthogonal, the SVD creates clusters of scatterers with different singular values. The ICA is then run with $N_{\rm{SV}}=44$ singular vectors in the matrix $V$, corresponding to singular values larger than 1$\%$ of the largest one, and with $N_{\rm{SV}}=33$ singular vectors, corresponding to those values larger than 10$\%$ of the largest one. Images are significantly improved, the cross being almost perfectly reconstructed with a blurring of about one pixel.

\begin{figure}[h!]
\begin{center}
    \includegraphics[scale=0.27]{Figs/cross-5000-f0-SV44.jpg}\hspace{-0.43cm}
    \includegraphics[scale=0.27]{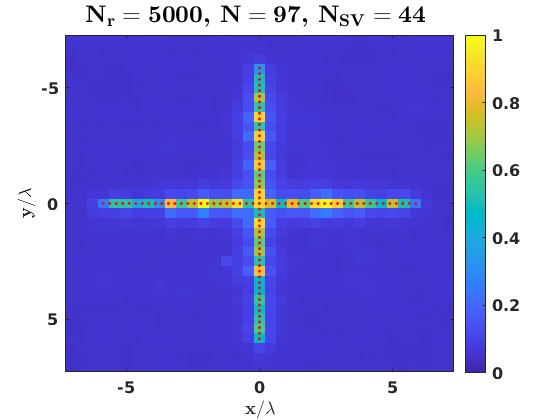}\hspace{-0.43cm}
    \includegraphics[scale=0.27]{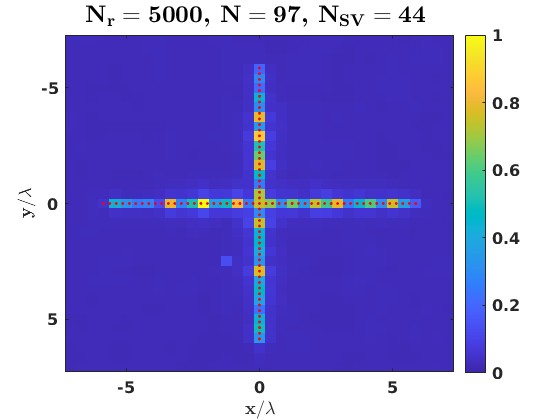}
  \caption{Effect of denoising for $N_{\rm{SV}}=44$ singular values and $N_r=5000$ realizations. Left: $\sigma=0$ (no denoising). Center $\sigma=0.4$. Right: $\sigma=0.6$. The case $\sigma=0.6$ yields almost pixel-size resolution with hardly any blurring, at the price of a lower contrast for some scatterers.}
  \label{fig:cross44N}
\end{center}
\end{figure}

In Figure \ref{fig:cross44N}, we investigate the effects of denoising on the image. As the denoising parameter $\sigma$ increases, blurring is reduced at the cost of a lower contrast for some well reconstructed scatterers. The case $\sigma=0.6$ yields almost pixel-size resolution with hardly any blurring.

\begin{figure}[h!]
\begin{center}
    \includegraphics[scale=0.27]{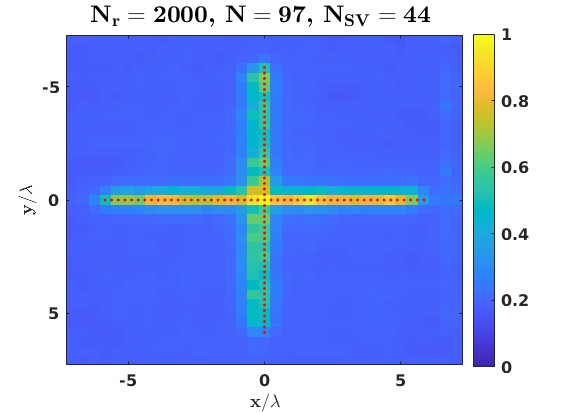}\hspace{-0.43cm}
    \includegraphics[scale=0.27]{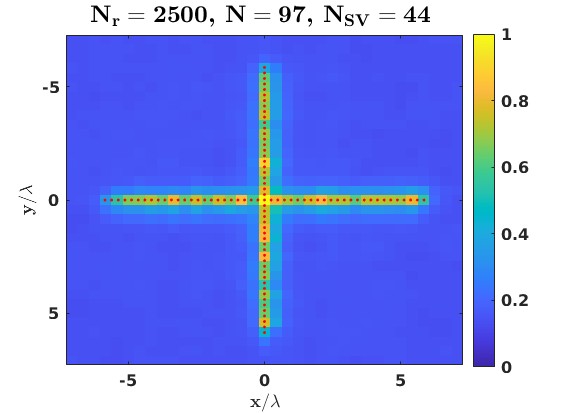}\hspace{-0.43cm}
    \includegraphics[scale=0.27]{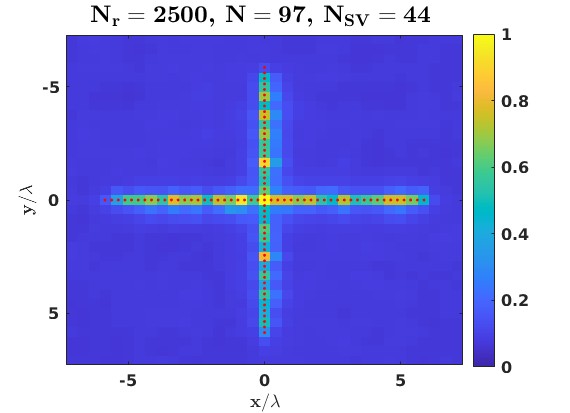}
   \caption{Imaging of a cross made with $N=97$ scatterers using only $N_{\rm{SV}}=44$ singular vectors (those larger than 1$\%$ of the greatest singular value). Left: $N_r=2000$ with $\sigma=0$ (no denoising). Center: $N_r=2500$ with $\sigma=0$. Right: $N_r=2500$ with $\sigma=0.3$. Blurring occurs for smaller $N_r$, which can be mitigated by denoising.}
  \label{fig:cross44N2}
\end{center}
\end{figure}

In Figure \ref{fig:cross44N2}, we decrease the number of illuminations to $N_r=2000$ and $N_r=2500$ when $N_{\rm{SV}}=44$ (left and center panels). A blurring effect is clearly observable compared to the case $N_r=5000$. This effect can be mitigated by setting $\sigma=0.3$ to obtain a high quality image (right panel) with about 50 times more illuminations than singular values included the reconstructions.



\begin{figure}[h!]
  \begin{center}
    \includegraphics[scale=0.27]{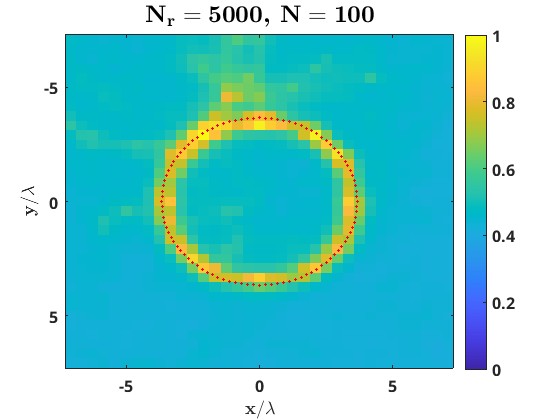}\hspace{-0.43cm}
  \includegraphics[scale=0.27]{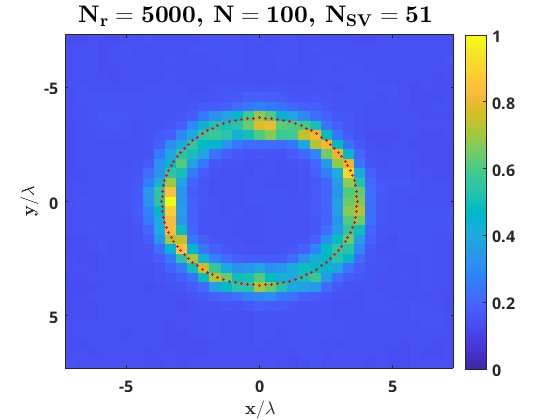}\hspace{-0.43cm}
  \includegraphics[scale=0.27]{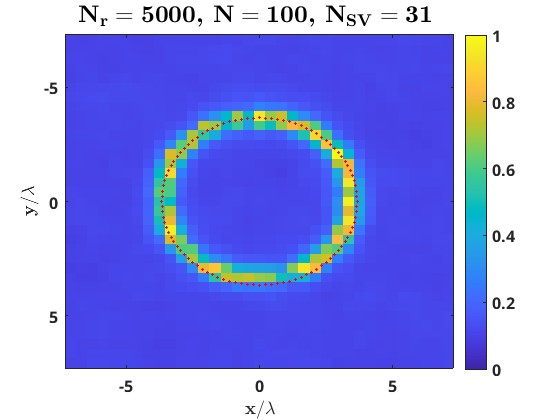}
   \caption{Imaging of a circle made with $N=100$ scatterers using only $N_{\rm{SV}}$ singular vectors for $N_r=5000$ realizations. Left: all singular vectors are used. Center: $N_{\rm{SV}}=51$ (those larger than 1$\%$ of the greatest singular value). Right: $N_{\rm{SV}}=31$ (those larger than 10$\%$ of the greatest singular value). The reconstructions improve when filtering out small singular values.}
  \label{fig:circ}
\end{center}
\end{figure}

We pursue our investigations with a circle of radius $15 h = 4.75 \lambda$ centered at the origin built with $N=100$ scatterers. We show in Figure \ref{fig:circ} reconstructions for $N_r=5000$ illuminations varying as before the number of singular values. With $N_{\rm{SV}}=51$, we keep singular values larger than 1$\%$ of the largest one, and with $N_{\rm{SV}}=31$ those larger than 10$\%$ of the largest one. As was observed for the cross, image quality is improved by removing small singular values, and the choice $N_{\rm{SV}}=31$ yields very good reconstructions.

\begin{figure}[h!]
\begin{center}
  \includegraphics[scale=0.27]{Figs/circle-5000-f0-SV31.jpg}\hspace{-0.43cm}
  \includegraphics[scale=0.27]{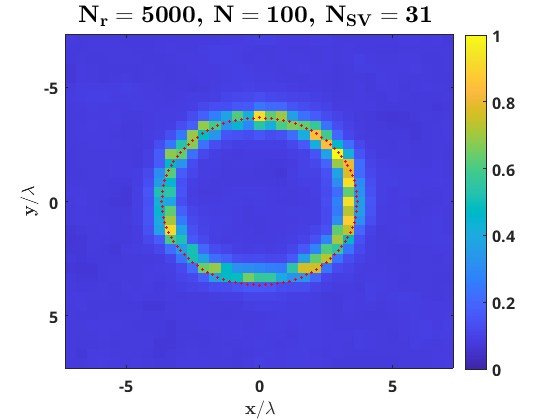}\hspace{-0.43cm}
  \caption{Imaging of a circle made with $N=100$ scatterers using only $N_{\rm{SV}}=31$ singular vectors (those larger than 1$\%$ of the greatest singular value). Left: $\sigma=0$ (no denoising). Right: $\sigma=0.3$. Denoising does not significantly change the image.}
  \label{fig:circ2}
\end{center}
\end{figure}

We add some denoising in Figure \ref{fig:circ2} and set $\sigma=0.3$, leading to minimal changes in the image.


\begin{figure}[h!]
\begin{center}
  \includegraphics[scale=0.265]{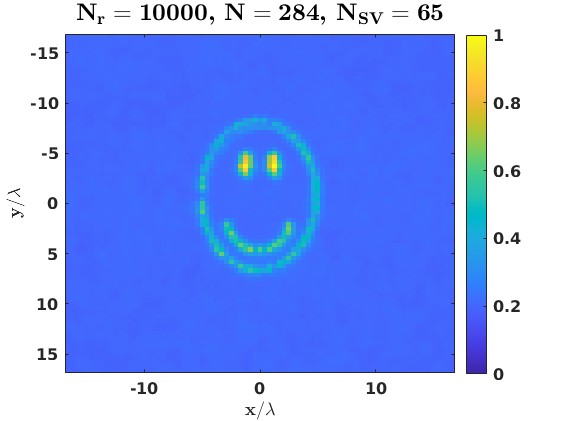}\hspace{-0.43cm}
    \includegraphics[scale=0.27]{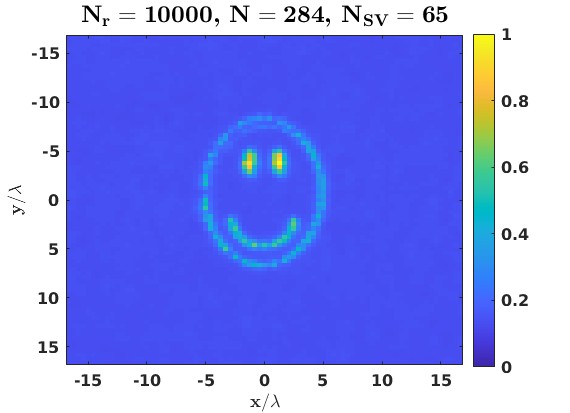}\hspace{-0.43cm}
      \includegraphics[scale=0.27]{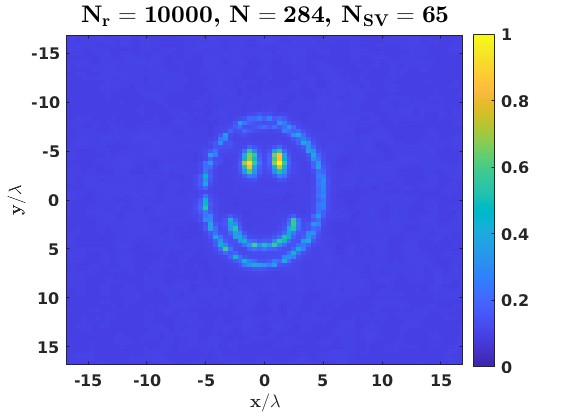}
    \caption{Imaging of a smiley made with $N=$ scatterers and $N_r=10000$ illuminations using only $N_{\rm{SV}}=65$ singular vectors (those larger than 10$\%$ of the greatest singular value). Left: $\sigma=0$ (no denoising). Center: $\sigma=0.2$. Right: $\sigma=0.3$. In this setting denoising has little impact.}
  \label{fig:smileF}
\end{center}
\end{figure}

\begin{figure}[h!]
\begin{center}
  \includegraphics[scale=0.27]{Figs/Smiley-10000-65-f0.jpg}
    \includegraphics[scale=0.277]{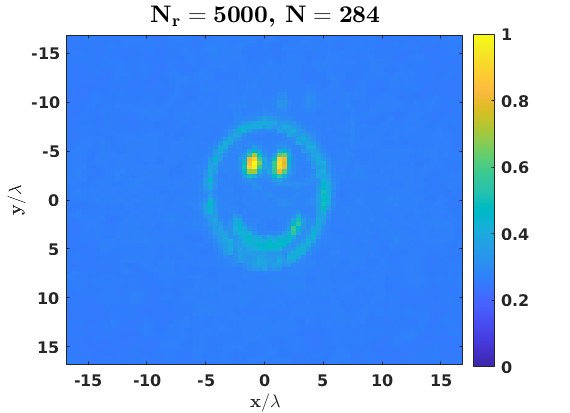}
     \caption{Imaging of a smiley made with $N=$ scatterers and $N_{\rm{SV}}=65$ singular vectors (those larger than 10$\%$ of the greatest singular value). Left: $N_r=10000$. Right: $N_r=5000$. In the last case, the image has more blurring and a weaker contrast.}
  \label{fig:smileF2}
\end{center}
\end{figure}

To conclude this section, we image a more complex object depicting a smiley built with $N=284$ scatterers. We represent in Figure \ref{fig:smileF} reconstructions obtained with $N_r=10000$ realizations, keeping $N_{\rm{SV}}=65$ singular vectors (those with singular values larger than 10$\%$ of the greatest singular value) and varying the denoising parameter. In this setting denoising has little impact. We decreased the number of illuminations to $N_r=5000$ in Figure \ref{fig:smileF2} and clearly observe poorer reconstructions.  

\subsection{Reconstructions beyond the memory effect window} \label{beyond}

\begin{figure}[h!]
\begin{center}
    \includegraphics[scale=0.27]{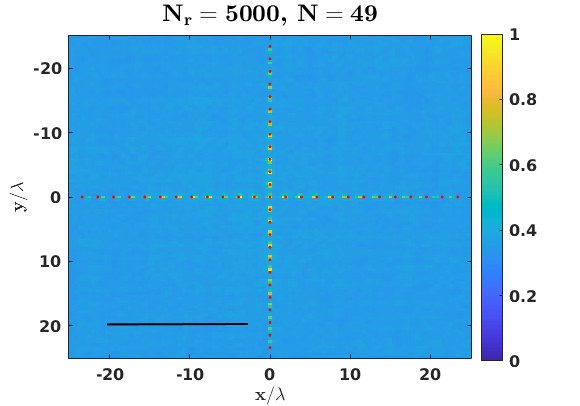}
    \includegraphics[scale=0.27]{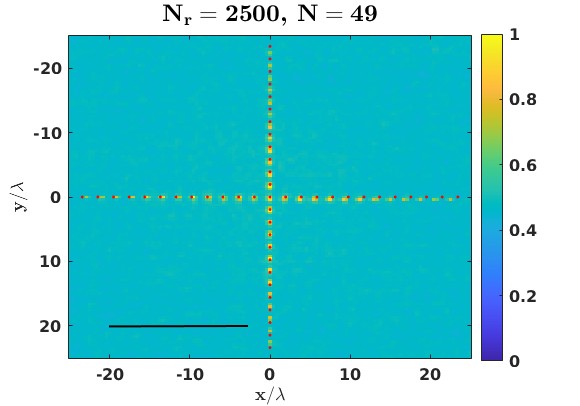}
  \caption{Imaging across multiple memory effect windows with $N=49$ uniformly placed scatterers. Theoretical memory length is $\ell_{\rm{me}}=11.31 \lambda$, while the numerical value is about $18 \lambda$. Left: $N_r=5000$ illuminations. Right $N_r=2500$. Scale bar represents the numerical memory length.}
  \label{fig:mem2}
\end{center}
\end{figure}

We display here reconstructions in a situation where the scatterers span multiple memory effect windows. We recall that the theoretical value is $\ell_{\rm{me}}=11.31 \lambda$, while the numerical value is about $18 \lambda$. In Figure \ref{fig:mem2}, we placed 49 scatterers uniformly horizontally and vertically across a total distance of about $50 \lambda$. Reconstructions are performed with $N_r=5000$ (right) and $N_r=2500$ (left) illuminations. The results are overall fairly accurate and comparable, with better signal-to-noise ratio in the case $N_r=5000$. We observe again the fact that scatterers situated in the periphery tend to be located with an error of a few pixels: small errors in the merge of the different images accumulate, resulting in the larger errors for remote scatterers. 

\subsection{Decreasing the number of illuminations} \label{decrease}

We recall that our strategy has been so far to exploit the non-Gaussianity of the fields $H_{\omega_0}^j$ to solve a blind source separation problem on the matrix $V^*$ coming from the SVD. This allows us to estimate the mixing matrix $A$ defined in \fref{defVA}, which in turn helps us separate the SHG fields $G_{2\omega_0}$. One inconvenience of the approach is the fairly large number of illuminations required for a suitable separation. The source separation problem cannot unfortunately be solved at the present time using the matrix $U$ defined in \fref{defU} since the SHG fields are Gaussian. If it were solvable, then pixels, or groups of pixels on the camera, can be seen as independent realizations of some random process and the source separation problem could be solved with far fewer random illuminations.

\begin{figure}[h!]
\begin{center}
  \includegraphics[scale=0.27]{Figs/Smiley-10000-65-f0.jpg}\hspace{-0.43cm}
  \includegraphics[scale=0.27]{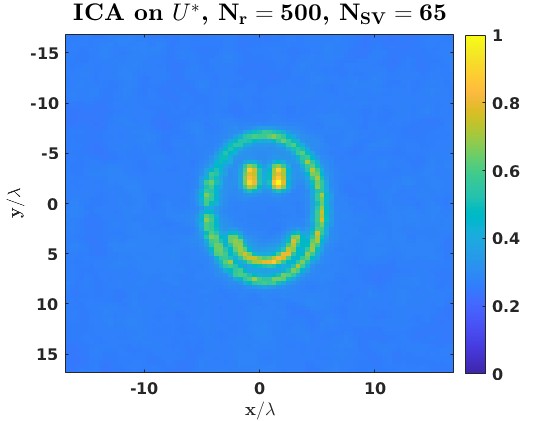}\hspace{-0.43cm}
      \includegraphics[scale=0.275]{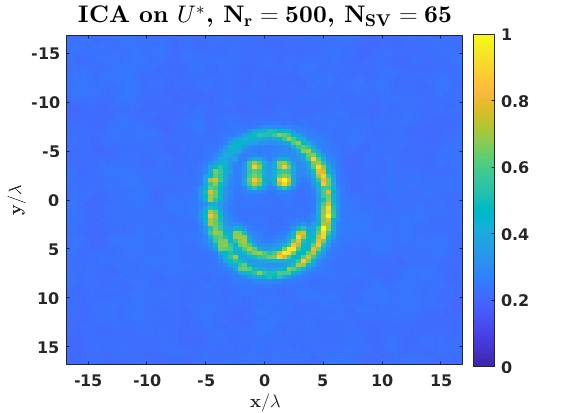}
     \caption{Reconstructions running ICA on $U^*$. Left: ICA on $V^*$ with $N_r=10000$ illuminations. Center: ICA on $U^*$ with $N_r=500$, denoising $\sigma=0$. Right: ICA on $U^*$ with $N_r=500$, denoising $\sigma=0.1$. Using $U^*$ yields quality images at a much lower cost.}
  \label{fig:U}
\end{center}
\end{figure}

In order to test the validity of this idea, we consider a hypothetical scenario where the SHG fields are not Gaussian. For this, we simply square the fields $G_{2\omega_0}$ to make them non-Gaussian, and solve the source separation problem with data $U^*$. We display the results in Figure \ref{fig:U}. The left image is the one obtained with $N_r=10000$ illuminations running the ICA on $V^*$ as before and keeping $N_{\rm{SV}}=65$ singular values. The center and right images are obtained with $N_r=500$ running the ICA on $U^*$, with denoising $\sigma=0$ and $\sigma=0.1$, respectively. The smiley is very well estimated with only 500 illuminations, to be contrasted with 10000 illuminations when using $V^*$.

We are currently working towards a method allowing us to separate Gaussian speckle fields emanating from point scatterers. Note that there are existing methods in the literature that were introduced for estimating the cosmic microwave background from measurements in the Planck space mission \cite{cardoso}, but those will likely not be directly applicable in our setting and will need to be modified. 

\section{Conclusion}

We developed in this work a method for imaging objects embedded in strongly diffusive environments. Our approach exploits independent component analysis techniques to extract individual Green's functions by improving the separation provided by the SVD and by resolving the speckle bond resonance issue that prevents any SVD separation in some configurations. The images are then obtained by using the memory effect and by estimating shifts in the Green's functions. This is accomplished  by solving a deconvolution problem with total variation regularization. The key quantity defining the resolution is the incoming speckle correlation length, which quantifies a lower limit under which the fields cannot be separated. This correlation length depends on the roughness of the underlying medium and on the imaging apparatus. It is equal to the diffraction-limited resolution length in our imaging setting, and we are therefore able to obtain images with diffraction-limited resolution provided the pixel size on the camera where measurements are made is sufficiently small. 

The method is robust in that it does not depend on the propagation model or on prior information about the illuminating patterns. It only relies on the fact that the incoming fields decorrelate at a given length and that the outgoing fields enjoy the memory effect. Our numerical simulations, performed in a realistic experimental setting, show the ability of the method to not only image discrete scatterers but also continuous objects. We also computed analytically various relevant correlation lengths and the memory length for our propagation model. 

At present, two main limitations constrain the method. The first one is that the random fields cannot be Gaussian in order to be separated. This lead us to consider an SHG imaging configuration where the fields are squares of Gaussian random fields. The second one is the fairly large number of illuminations required to obtain good images. We found numerically that if $N$ is the number of scatterers (or the number of singular values) to be imaged, we need between fifty and a hundred times more illuminations when $N \leq 100$. We are currently working on a method allowing us to treat both issues at once: separating Gaussian fields would not only extend the applicability of our imaging strategy to non-SHG settings, but would also reduce the number of illuminations needed, as explained in Section \ref{decrease}, as groups of pixels can be seen as independent realizations on the camera. 

Our method will next be tested on experimental data that are presently under collection.

    \begin{appendix}
    
      \section{Memory effect} \label{app:memory}
      We investigate in this section the memory effect for our propagation model, and establish analytical expressions for the memory length $\ell_{\rm{me}}$ and the scaling factor $\alpha_{\rm{me}}$. We consider one Green's function $G_{2 \omega_0}$ originating from the point $\bx_i$, and one from the origin. Our goal is to obtain conditions on $|\bx_i|$ under which these two Green's functions are approximately shifted versions of each other and to identify the shift.
      
      We start by rewriting \fref{bG1}-\fref{bG2} in terms of $G_{2 \omega_0}$: with $\calT_j(\by)=P(\by) e^{i k_2  S_j(\by)}$, for $j=0,1,$ and $k_2=2k$, we have
      \bee
      G_{2 \omega_0}(\bx,2f;\bx_i,0)&=&e^{i \frac{k_2 L}{2 f^2} |\bx|^2} \int_{\Rm^2} e^{-i \frac{k_2}{f} \bx \cdot \by} G_{2 \omega_0}(\by,L;\bx_i,0) \calT_1 (\by)d\by\\
      G_{2 \omega_0}(\bx,L;\bx_i,0)&=&\int_{\Rm^2} e^{i \frac{k_2}{2L_1} |\bx-\by|^2} e^{i \frac{k_2}{2L_0} |\by-\bx_i|^2} \calT_0 (\by)d\by.\\
      \eee
      Introducing $\bv_0=-L_1\bx_i/L_0$, we recast $G_{2 \omega_0}(\bx,L;\bx_i,0)$ as
\bee
G_{2 \omega_0}(\bx,L;\bx_i,0)&=&e^{i \frac{k_2}{2L_1}|\bx|^2} e^{i \frac{k_2}{2L_0} |\bx_i|^2}\int_{\Rm^2} e^{i \frac{k_2 L}{2L_0L_1} |\by|^2} e^{-i \frac{k_2}{L_1} y \cdot (\bx-\bv_0)}  \calT_0 (\by)d\by\\
&=&e^{i \frac{k_2}{2L_1}|\bx|^2} e^{i \frac{k_2}{2L_0} |\bx_i|^2} F_0(\bx-\bv_0)
\eee
where
$$
F_0(\bx)=\int_{\Rm^2} e^{i \frac{k_2 L}{2L_0L_1} |\by|^2} e^{-i \frac{k_2}{L_1} \by \cdot \bx} \calT_0 (\by)d\by.
$$
Direct algebra then shows that
$$
G_{2 \omega_0}(\bx,2 f;\bx_i,0)=e^{i \frac{k_2 L}{2 f^2} |\bx|^2} e^{i \frac{k_2}{2L_0} |\bx_i|^2}e^{-i \frac{k_2}{f} \bx \cdot \bv_0}e^{i \frac{k_2}{2L_1}|\bv_0|^2}F_1(\bx+\bv_1,\bx_i),
$$
where $\bv_1=f \bx_i /L_0$ and
$$
F_1(\bx,\bx_i)=\int_{\Rm^2} e^{-i \frac{k_2}{f} \bx\cdot \by} e^{i \frac{k_2}{2L_1}|\by|^2}F_0(\by) \calT_1 (\by+\bv_0)d\by.
$$
Ignoring the complex exponentials in $G_{2 \omega_0}(\bx,2 f;\bx_i,0)$, we observe the memory effect provided $|F_1(\bx+\bv_1,\bx_i)| \simeq |F_1(\bx+\bv_1,\bzero)|$, and this shows that the scaling factor is $\alpha_{\rm{me}}=f/L_0$. Our main task here is therefore is to obtain a condition on $|\bx_i|$ under which indeed $|F_1(\bx+\bv_1,\bx_i)| \simeq |F_1(\bx+\bv_1,\bzero)|$. For this, we calculate the correlation between these two quantities and use the fact that $G_{2 \omega_0}(\bx,2 z_s;\bx_i,0)$ is in a speckle regime. Hence, $F_1 $ is approximately a circular complex Gaussian field and we have then, according to Isserlis theorem,
\begin{align} \nonumber
  \E\Big\{(|F_1(\bx;\bx_i)|^2&-\E\{|F_1(\bx;\bx_i)|^2\} )(|F_1(\bx;\bzero)|^2-\E\{|F_1(\bx;\bzero)|^2\})\Big\} \\
  &\simeq \; |\E\{F_1(\bx;\bx_i) F_1(\bx;\bzero)\}|^2+|\E\{F_1(\bx;\bx_i) F_1^*(\bx;\bzero)\}|^2,\label{gaussF}
\end{align}
where the expectation is taken with respect to realizations of the random fields $V_0$ and $V_1$. We begin with the last term in the sum above, which is the most important, and will use  the notation
$$
\E\{\calT_j(\bx)\calT_j^*(\by)\}=P(\bx)P^*(\by)\calC_j((\bx-\by)/\ell_{c,j}), \qquad j=0,1.
$$
The phase screen correlation lengths $\ell_{c,0}$ and  $\ell_{c,1}$ are defined in Section \ref{model}. We have
\bee
\E\{F_0(\bx)F_0^*(\bx')\}&=&\int_{\Rm^4} e^{i \frac{k_2 L}{2L_0L_1} (|\by|^2-|\by'|^2)} e^{-i \frac{k_2}{L_1} \by \cdot \bx}e^{i \frac{k_2}{L_1} \by' \cdot \bx'} P(\by)P^*(\by')\calC_0\left(\frac{\by-\by'}{\ell_{c,0}}\right)d\by d\by'\\
&=&\int_{\Rm^4} e^{-i \frac{k_2 L}{2L_0L_1}|\ell_{c,0} \by'|^2}e^{-i \frac{k_2}{L_1} \by \cdot [(\bx-\bx')-\ell_{c,0} L \by'/L_0]}e^{i \frac{k_2}{L_1} \ell_{c,0} \by' \cdot \bx'} \\
&& \hspace{3cm} \times P(\by)P^*(\by+\ell_{c,0} \by')\calC_0(-\by')d\by d\by',
\eee
where the last equality is up to irrelevant multiplicative constants. Neglecting the finite size of the random screen by setting $P(\by)=1$, the integral in $\by$ yields a Dirac delta and the expression reduces to
\bea \label{EF0}
\E\{F_0(\bx)F_0^*(\bx')\}&\simeq &e^{-i \frac{k_2 L_0}{2L_1 L} |\bx-\bx'|^2}e^{-i \frac{k_2 L_0}{L L_1} (\bx-\bx')\cdot \bx'} \calC_0((\bx'-\bx)L_0/L \ell_{c,0}).
\eea
We now turn to $F_1$ and find, since the random fields in $\calT_0$ and $\calT_1$ are independent,
\begin{align*}
  \E\{F_1(\bx,&\bx_i)F_1^*(\bx,\bzero)\}\\
  &=\int_{\Rm^4} e^{-i \frac{k_2}{f} \bx\cdot (\by-\by')} e^{i \frac{k_2}{2L_1}(|\by|^2-|\by'|^2)}\E\{F_0(\by)F_0^*(\by')\} \\
&\qquad \times P(\by+\bv_0)P^*(\by')\calC_1((\by+\bv_0-\by')/\ell_{c,1})d\by d\by'\\
&=\int_{\Rm^4} e^{i k_2 \ell_{c,1} (\bx/f-(\by+\bv_0)/L_1)\cdot \by'} e^{-i \frac{k_2}{L_1} \by \cdot \bv_0} e^{-i \frac{k_2}{2L_1} |\ell_{c,1} \by'|^2}\\
&\qquad \times \E\{F_0(\by)F_0^*(\by+\bv_0+\ell_{c,1} \by')\} P(\by+\bv_0)P^*(\ell_{c,1} \by'+\by+\bv_0) \calC_1(-\by')d\by d\by',
\end{align*}
where the last equality is up to irrelevant factors. 
Using \fref{EF0}, we have
\begin{align*}
  \E\{F_0(\by)F_0^*(\by+\bv_0+\ell_{c,1} \by')\}=e^{-i \frac{k_2 L_0}{2LL_1}|\bv_0+\ell_{c,1} \by'|^2}&e^{i \frac{k_2 L_0}{L L_1} (\bv_0+\ell_{c,1} \by')\cdot (\by+\bv_0+\ell_{c,1} \by')} \\
  &\times\calC_0((\bv_0+\ell_{c,1} \by')L_0/L \ell_{c,0}).
\end{align*}
Neglecting once more the finite size of $P$ by setting $P(\by)=1$, the $\by$ integral in $\E\{F_1 F_1^*\}$ produces a Dirac delta that forces $\by'= \bv_0 / \ell_{c,1}$. This finally gives, up to unimportant factors,
$$
|\E\{F_1(\bx,\bx_i)F_1^*(\bx,\bzero)\}|\simeq \calC_0(\bzero) \calC_1(\bx_i/\ell_{\rm{me}}), \qquad \ell_{\rm{me}}=\frac{L_0 \ell_{c,1}}{L_1}.
$$

We conclude by observing that similar calculations show that the first term in \fref{gaussF} is negligible compared to the last one as it involves terms of the form $\E\{\calT_j(\bx)\calT_j(\by)\}$ that are small compared to $\E\{\calT_j(\bx)\calT^*_j(\by)\}$. Besides, the average $\E\{|F_1(\bx;\bx_i)|^2\}$ is approximately constant w.r.t. $\bx_i$ since it only depends on $\bx_i$ through slow varying terms of the form $P(\by+\bv_0)$. As a consequence, the quantity $\E\{|F_1(\bx;\bx_i)|^2 |F_1(\bx;\bzero)|^2\}$ varies w.r.t. $\bx_i$ at the scale defined by $|\E\{F_1(\bx,\bx_i)F_1^*(\bx,\bzero)\}|^2$, which is $\ell_{\rm{me}}$.


\section{Field correlations} \label{app:corr}
We characterize in this section the spatial correlation length of the speckle, which is the main indicator of the separability of the fields. We start with the incoming field.

\paragraph{The forward field.} 
We first recast the fundamental field $U_{\omega}^{(1)}(\bx,0)$ defined in \fref{UL} as 
\bee
U_{\omega_0}^{(1)}(\bx,0) &=& e^{ikL_0} \int_{\Rm^2}  e^{i \frac{k  L}{2f^2} |\bv|^2} O(\bx,\bv) \calT_{\rm{SLM}}(\bv)d\bv
\eee
where
\bee
O(\bx,\bv)&=&\int_{\Rm^4} e^{i \frac{k}{2L_0} |\bx-\by|^2} e^{i \frac{k}{2L_1} |\by-\bz|^2}  e^{-i \frac{k}{f} \bz \cdot \bv} \calT_{0}(\by)\calT_{1}(\bz)d\by d\bz.
\eee
With the notations
\bee
\E\{\calT_{\rm{SLM}}(\bx)\calT_{\rm{SLM}}^*(\by)\}&=&P_{\rm{SLM}}(\bx)P_{\rm{SLM}}^*(\by)\calC_{\rm{SLM}}((\bx-\by)/l_{c,{\rm{SLM}}}), \qquad j=0,1,\\
\E\{O(\bx,\bv)O^*(\bx',\bv')\}&=&E(\bv,\bv',\bx,\bx'),
\eee
where $l_{c,{\rm{SLM}}}$ is introduced in Section \ref{model}, we find
\bea  \label{EU}
\E\{U_{\omega_0}^{(1)}(\bx,0)(U_{\omega_0}^{(1)})^*(\bx',0) \}&=& \int_{\Rm^4}  e^{-i \frac{k  L}{f^2}\bv \cdot \bv'} e^{-\frac{i kL}{2f^2} |\bv'|^2 }E(\bv,\bv+\bv',\bx,\bx') \\
&&\; \times P_{\rm{SLM}}(\bv)P^*_{\rm{SLM}}(\bv+\bv')\calC_{\rm{SLM}}(-\bv'/l_{c,\rm{SLM}})d\bv d\bv'. \nonumber
\eea
Tedious but not difficult calculations show that
\begin{align*}
E(\bv,&\bv+\bv',\bx,\bx') 
  =e^{\frac{i kL}{2f^2} |\bv'|^2 } e^{-i \frac{k}{f} (\bx-\bx')\cdot \bv} e^{i \frac{k L}{f^2} \bv'\cdot \bv} e^{i \frac{k}{f} \bx'\cdot \bv'}
  \\
  &\hspace{3cm}\calC_0\big([\bv' L_0/f+(\bx'-\bx)]/l_{c,0}\big)\calC_1\big([\bv'L_0/f+(\bx'-\bx)]L/L_0l_{c,1}\big),
\end{align*}
where the last equality is up to irrelevant multiplicative constants. We have now all needed to conclude. When  $P_{\rm{SLM}}$ is the indicator function of a disk of radius $ \mathrm{NA} f$ (we recall that $ \mathrm{NA}$ is the numerical aperture of the microscope and $f$ its focal) and $l_{c,SLM}\ll  \mathrm{NA} f$, (in practice $l_{c,SLM}$ is of the order the $\mu$m while $f$ is of the order of the mm and $ \mathrm{NA}$ is between 0.2 and 1), we can neglect the $\bv'$ dependence in $P^*_{\rm{SLM}}(\bv+\bv')$ and approximate it by $P^*_{\rm{SLM}}(\bv)$. The $\bv$ integral in \fref{EU} then yields the function
$$
Q( \mathrm{NA} (\bx-\bx')/\lambda)=\int_{|\bv| \leq 1} e^{-i \frac{2 \pi  \mathrm{NA}}{\lambda} (\bx-\bx')\cdot \bv} d\bv,
$$
which the usual free space Point Spread Function that exhibits the diffraction limited resolution $\lambda/2  \mathrm{NA}$. 

Expression \fref{EU} consequently reduces to
\begin{align*}
  &\E\{U_{\omega_0}^{(1)}(\bx,0)(U_{\omega_0}^{(1)})^*(\bx',0) \}=Q( \mathrm{NA} (\bx-\bx')/\lambda)\\[1mm]
  &\times \int_{\Rm^2}  e^{i \frac{k}{f} \bx'\cdot \bv'}\calC_{\rm{SLM}}\left(-\frac{\bv'}{l_{c,\rm{SLM}}}\right)\calC_0\left(\frac{\bv' L_0/f+(\bx'-\bx)}{l_{c,0}}\right)\calC_1\left(\frac{[\bv'L_0/f+(\bx'-\bx)]L}{L_0l_{c,1}}\right)d\bv'.
\end{align*}
Assuming the function $\calC_{\rm{SLM}}$ is integrable, the Lebesgue dominated convergence theorem shows that the above quantity converges to zero as $|\bx-\bx'| \gg \min(l_{c,0},l_{c,1}L_0/L)$.
The conclusion is therefore that the speckle correlation length $\ell_{s,\rm{in}}$ of the incoming field is
$$
\ell_{s,\rm{in}}=\min\left( \lambda/2 \mathrm{NA},l_{c,0},L_0l_{c,1}/L\right).
$$

\paragraph{The backward field.} The situation is similar to the one of the last section, with the difference that $k$ is replaced by $2 k$ and that the aperture $P_{\rm{SLM}}$ is replaced by the indicator function of the domain where measurements are made. If that domain has a diameter $D$, then the correlation length for the SHG speckle is
$$
\ell_{s,\rm{out}}=\min\left( \lambda f/2D,\ell_{c,0},L_0\ell_{c,1}/L\right).
$$

\section{The Lindeberg condition} \label{app:linde}
We recall that
$$
X_\ell(\mu)=\int_{S_1}e^{-i 2 \pi \mu (\by_\ell+\by)_1}  e^{i2kS_1(\ell_{c,1}(\by_\ell+\by))} \calG_{2\omega_0}(\ell_{c,1}(\by_\ell+\by),L;\bzero,0)d\by
$$
and that the Lindeberg condition reads, for a collection of real-valued random variables $Y_k$,
\be \label{linde}
\frac{1}{s_n^2} \sum_{\ell=1}^n \mathbb{E}\left\{ Y_{\ell}^2 \mathbf{1}_{\{|Y_{\ell}| > \epsilon s_n\}} \right\} \to 0 \quad \text{as } n \to \infty,
\ee
for every \( \epsilon > 0 \), and where \( s_n^2 = \sum_{\ell=1}^n \mathrm{Var}(Y_{\ell}) \). Since $X_\ell(\mu)$ is complex-valued, we consider the real-valued random variable $Z_\ell=a \Re X_\ell(\mu) + b \Im X_\ell(\mu)$ for arbitrary $a,b \in \Rm$. As $|\E\{X_\ell(\mu)^2\}| \ll \E\{|X_\ell(\mu)|^2\}$ and $\E\{X_\ell(\mu)\} \simeq 0$, the variance of $Z_\ell$ is dominated by $\E\{|X_\ell(\mu)|^2\}$ and it is enough to check \fref{linde} with $Y_\ell=|X_\ell(\mu)|$ and $s_n^2 = \sum_{\ell=1}^n \E\{|X_\ell(\mu)|^2\}$.

\paragraph{Step 1.} The first step is to show that $s_n^2 \geq C n$ (for us $n=4M^2$) for some constant $C>0$. For this, adapting the calculations of Appendix \ref{app:memory}, we find
\begin{align*}
\E\big\{\calG_{2\omega_0}(\bx,L;\bzero,0) &\calG_{2\omega_0}(\bx',L;\bzero,0)^*\big\}\\
&\simeq e^{\frac{i k_2}{2L_1} (|\bx|^2-|\bx'|^2)} e^{-i \frac{k_2 L_0}{2L L_1} |\bx-\bx'|^2}e^{-i \frac{k_2 L_0}{L L_1} (\bx-\bx')\cdot \bx'} \calC_0\left(\frac{(\bx'-\bx)L_0}{L \ell_{c,0}}\right),
\end{align*}
so that
\begin{align*}
\E\big\{\calG_{2\omega_0}&(\ell_{c,1}(\by_\ell+\by),L;\bzero,0) \calG_{2\omega_0}(\ell_{c,1}(\by_\ell+\by'),L;\bzero,0)^*\big\}\\
                     &\simeq e^{\frac{i k_2\ell_{c,1}^2}{2L_1} (\by-\by')\cdot (2\by_\ell+\by+\by')} e^{-i \frac{k_2 L_0 \ell_{c,1}^2}{2L_1 L} |\by-\by'|^2}e^{-i \frac{k_2 L_0 \ell_{c,1}^2}{L L_1} (\by-\by')\cdot (\by_\ell+\by')}\calC_0\left(\frac{(\by'-\by)\ell_{c,1} L_0}{L \ell_{c,0}}\right).
\end{align*}
This yields
\begin{align*}
  \E\{|X_\ell(\mu)|^2\} &\simeq \int_{S_1 \times S_1} T_\ell(\by,\by') T(\by,\by') d\by d\by'
\end{align*}
where
\begin{align*}
  T_\ell(\by,\by') &=e^{\frac{i k_2 \ell_{c,1}^2}{L} (\by-\by')\cdot \by_\ell}\\
      T(\by,\by')&=e^{-i 2 \pi \mu (y_1-y'_1)}e^{\frac{i k_2\ell_{c,1}^2}{2L} (|\by|^2-|\by'|^2)} \calC_1(\by'-\by)\calC_0\left(\frac{(\by'-\by)\ell_{c,1} L_0}{L \ell_{c,0}}\right).
\end{align*}
We recast the centers $\by_\ell$ of the squares in cartesian coordinates as $\by_\ell=(m+\frac{1}{2},p+\frac{1}{2})$ and $T_\ell$ accordingly as $T_{m,p}$. It follows that
\begin{align*}
\sum_{\ell=1}^{4 M^2} &T_\ell(\by,\by')=\sum_{m=-M}^{M-1}\sum_{p=-M}^{M-1}  T_{m,p}(\by,\by')\\
&=e^{\frac{i \gamma (y_1-y_1')}{2}}e^{\frac{i \gamma (y_2-y_2')}{2}}\left(D(\gamma (y_1-y_1'))-e^{i M \gamma (y_1-y_1')}\right)\left(D(\gamma (y_2-y_2'))-e^{i M \gamma (y_2-y_2')}\right),
\end{align*}
where $\gamma=\frac{k_2\ell_{c,1}^2}{L}$ and $D$ is the Dirichlet kernel $D(x)=\sin((M+1/2)x)/\sin(x/2)$. In order to conclude, we remark that $D(0)=2M+1$, and that $\gamma \ll 1$ (in our simulations we have $M=177$ and $\gamma=0.06$). Since $|y_j-y_j'|\leq 2$, there is then a constant $C$, independent of $M$, such that $D(\gamma (y_1-y_1')) \geq C M$, and since $M \gg 1$, the sum of the $T_\ell$ is approximately equal to
\bee
\sum_{\ell=1}^{4 M^2} T_\ell(\by,\by')&\simeq &e^{\frac{i \gamma (y_1-y_1')}{2}}e^{\frac{i \gamma (y_2-y_2')}{2}} D(\gamma (y_1-y_1')) D(\gamma (y_2-y_2')).
\eee
Moreover, $\E\{|X_\ell(\mu)|^2\}$ is real-valued as well as $\calC_0$ and $\calC_1$, and we have
\begin{align*}
  \sum_{\ell=1}^{4 M^2}  \E\{|X_\ell(\mu)|^2\} &\simeq \int_{S_1 \times S_1} \cos (\theta(\by,\by')) T'(\by,\by')d\by d\by'
\end{align*}

where
\bee
\theta(\by,\by')&=&\frac{\gamma}{2}\left(y_1-y_1'+y_2-y_2'+|\by|^2-|\by'|^2\right)+2 \pi \mu (y_1'-y_1)\\
T'(\by,\by')&=&D(\gamma (y_1-y_1'))D(\gamma (y_2-y_2'))\calC_1(\by'-\by)\calC_0\left(\frac{(\by'-\by)\ell_{c,1} L_0}{L \ell_{c,0}}\right).
\eee
Using that $\gamma \ll 1$ and setting e.g. $4 \pi \mu \leq \pi /4$, there is a constant $C'$ such that $\cos (\theta(\by,\by')) \geq C'$ for $\by \times \by' \in S_1 \times S_1$. Hence, finally,
\be \label{boundbelow}
\sum_{\ell=1}^{4 M^2}  \E\{|X_\ell(\mu)|^2\}  \geq  C' \int_{S_1 \times S_1} T'(\by,\by')d\by d\by' \geq C'' M^2,
\ee
for some other constant $C''$ independent of $M$. This bounds the $s_n$ term from below in \fref{linde}. We tackle now the numerator in \fref{linde}. 
\paragraph{Step 2.} Calculations using the Gaussian character of the potential $V_0$ show that there is a constant $C$ independent of $M$ such that
\be \label{4B}
\E\{ |X_\ell(\mu)|^4\} \leq C.
  \ee
  Using the Cauchy-Schwarz inequality and \fref{4B}, we find
  \bee
  \sum_{\ell=1}^n \mathbb{E}\left\{ Y_{\ell}^2 \mathbf{1}_{\{|Y_{\ell}| > \epsilon s_n\}} \right\} &\leq& C^{1/2}\sum_{\ell=1}^n \left(\mathbb{P}(|Y_{\ell}| > \epsilon s_n)\right)^{1/2} \leq C^{1/2}\sum_{\ell=1}^n \left(\frac{\E\{ Y_\ell^4\}}{(\epsilon s_n)^4}\right)^{1/2}\\
  &\leq& \frac{C n}{(\epsilon s_n)^2},
  \eee
  which, together with \fref{boundbelow}, shows that the Lindeberg condition \fref{linde} is satisfied.
\end{appendix}


\footnotesize{
 \bibliographystyle{siam}
  \bibliography{bibliography.bib} }
 \end{document}